\documentclass[10pt,journal,compsoc]{IEEEtran}

\usepackage{cite}

\usepackage{setspace}
\usepackage{epsf}\epsfverbosetrue
\usepackage{graphics,epsfig}
\usepackage{graphicx,epsfig}
\usepackage{epsfig}
\usepackage{multirow}
\usepackage{alltt}
\usepackage{subfigure}
\usepackage{tcolorbox}
\usepackage{float}
\usepackage{url}
\usepackage{graphicx}
\usepackage{verbatim} 
\usepackage{footnote} 
\usepackage{latexsym} 
\usepackage{sidecap} 
\usepackage{wrapfig}
\usepackage{eso-pic}
\usepackage{fix-cm}
\usepackage{algorithm}
\usepackage{algpseudocode}
\usepackage{color}
\usepackage{wrapfig}
\usepackage{multicol}

\usepackage{layout}

\usepackage{amsmath}
\usepackage[
  locale = DE 
]{siunitx}

\usepackage{textcomp}
\usepackage{multirow}
\usepackage{mathtools}
\usepackage{soul} 
\usepackage{amsmath} 
\usepackage{verbatim}
\usepackage{breqn}
\usepackage{csvsimple} 
\usepackage{array}
\usepackage{subfig}
\usepackage[normalem]{ulem}
\usepackage{booktabs}
\newcolumntype{P}[1]{>{\RaggedRight\arraybackslash}p{#1}}
\newcommand{\tabitem}{\textbullet~~}

\usepackage{mathtools}
\DeclarePairedDelimiter\ceil{\lceil}{\rceil}
\DeclarePairedDelimiter\floor{\lfloor}{\rfloor}

\usepackage{pifont}

\newcommand{\simi}[1]{\textcolor{black}{#1}}

\newcommand{\fin}[1]{\textcolor{black}{#1}}

\newcommand{\modify}[1]{\textcolor{black}{#1}}
\newcommand{\tkde}[1]{\textcolor{black}{#1}}

\newcommand{\tkderev}[1]{\textcolor{black}{#1}}

\newcommand{\mubtkde}[1]{\textcolor{black}{#1}}

\newtheorem{theorem}{\textbf{Theorem}}
\newtheorem{lemma}{\textbf{Lemma}}

\begin{document}

\title{Differentially Private Demand Side Management for Incentivized Dynamic Pricing in Smart Grid\footnotesize$^{^{^{^{^{^1}}}}}$}

\author{\IEEEauthorblockN{Muneeb Ul Hassan\IEEEauthorrefmark{1}, Mubashir Husain Rehmani\IEEEauthorrefmark{4}, Jia Tina Du\IEEEauthorrefmark{6}, Jinjun Chen\IEEEauthorrefmark{1}\\}
\IEEEauthorblockA{\IEEEauthorrefmark{1}Swinburne University of Technology, Hawthorn VIC 3122, Australia\\ \IEEEauthorrefmark{4} Munster Technological University (MTU), Ireland\\ \IEEEauthorrefmark{6} University of South Australia, Adelaide, Australia}
}


\IEEEtitleabstractindextext{%
\begin{abstract}
In order \tkde{to efficiently provide demand side management (DSM) in smart grid, carrying out pricing on the basis of real-time energy usage} is considered to be the most vital tool because it is directly linked with the finances associated with smart meters. \tkderev{Hence, every smart meter user wants to pay the minimum possible amount along with getting maximum benefits.} \mubtkde{In this context, usage} based dynamic pricing strategies of DSM plays their role and provide users with specific incentives that help shaping their load curve according to the forecasted load. However, these reported real-time values can leak privacy of smart meter users, which can lead to serious consequences such as spying, etc. \tkderev{Moreover, most dynamic pricing algorithms charge all users equally irrespective of their contribution in causing peak factor.} Therefore, in this paper, we propose a modified usage based dynamic pricing  mechanism that only charges the users responsible for causing peak factor. We further integrate the concept of differential privacy to protect the privacy of real-time \mubtkde{smart metering data. To calculate accurate billing, we also propose a noise adjustment method.} Finally, we propose \textbf{D}emand \textbf{R}esponse enhancing \textbf{D}ifferential \textbf{P}ricing (DRDP) strategy that effectively enhances demand response along with providing dynamic pricing to smart meter users. \tkderev{We also carry out theoretical analysis for differential privacy guarantees and for cooperative state probability to analyze behavior of cooperative smart meters.} The performance evaluation of DRDP strategy at various privacy parameters show that the proposed strategy outperforms previous mechanisms in terms of dynamic pricing and privacy preservation.
\end{abstract}

}

\maketitle

\footnotetext[1]{A preliminary version has been published by 2020 IEEE International Conference on Communications (ICC 2020), June, 2020, Dublin, Ireland entitled ``Differentially Private Dynamic Pricing for Efficient Demand Response in Smart Grid”. \newline This paper is partly supported by Australian Research Council (ARC) projects DP190101893, DP170100136, LP180100758.}

\section{Introduction}

 \tkderev{Modern day smart homes are equipped with smart meters which send their real-time energy usage values to smart grid utility in order to carry out plenty of tasks such as demand side management (DSM), load forecasting, etc}~\cite{newint01}. \tkderev{This real-time energy usage data is used to formulate strategies that help shape the load curves and carry out efficient load utilization (ELU).} ELU is a method of shaping smart homes \mubtkde{energy usage in such a way} that it equates with the possible energy supply in the specific time instant~\cite{newint02}.  In order to do so, demand side management (DSM) \mubtkde{strategies are proposed,} which shape the load curves by providing interesting and timely incentives to participating smart homes~\cite{tkdenew14}. Similarly, almost all DSM strategies (also known as demand response (DR) strategies) have a common goal, which is to motivate smart homes users to use minimum energy during peak load times and to shift surplus energy usage to off-peak times (e.g., using washing machine in off-peak hours)~\cite{tkderef01}. \\
\tkderev{Till now, plenty of DR models \mubtkde{have been proposed,} for example, control mechanism models, offered motivation models, and decision variable models. Among them, offered motivation models are the most popular ones which are further categorized into price based models and incentive based models~\cite{intref01, tkdenew03}.} In offered motivation models, dynamic pricing dominates other models because it provides users the maximum control to get incentivized. In dynamic pricing mechanisms, users are charged with respect to the rate being devised by grid utility, so users can orient their usage at the time of low rates and use heavy appliances at off-peak hours. This model is somehow beneficial, but it has a major flaw that \textit{what if all smart homes start using their heavy appliances \tkde{at once at the time} of low pricing hours?} In fact, if this happens, then the low-pricing hours can cause a shortage of electricity as it was not predicted during load forecasting and strategy to overcome this sudden shortage was not developed. In order to overcome this, researchers came up with the idea of dynamic peak hours, which means that peak hours are not fixed and can vary with respect to energy usage within a specific area. \mubtkde{This is also known as dynamic peak factor model.} \tkderev{For example, if energy exceeds a specific peak value, then the peak-hour is in place and smart homes will be charged peak hour price~\cite{litref01, tkdenew04}.}\\
\tkderev{Overall, this dynamic peak factor model is well-suited to meet the demands of load forecasting, but on the \tkde{other hand, it has two} major issues from the perspective of smart homes. Firstly, it also charges the same high peak factor price to smart homes which are not responsible for causing that peak-hour.} Secondly, the collection of fine-grained data of smart homes for load forecasting and for peak hour determination raises serious threats to privacy leakage of smart home users. E.g., this real-time data can further be used to carry out various malicious activities such as forgery, routine tracking, etc. Similarly, this data can also be fed-up to non-intrusive load monitoring (NILM) \tkde{mechanisms, these techniques predict the usage of a specific household appliance (such as toaster, washing machine, etc.) at a specific slot of time}~\cite{tkderef02}. These NILM mechanisms can even find out any faulty appliance and can estimate its possible day of breakdown, which can further be used to carry out targeted advertisement~\cite{litref06}. Therefore, a mechanism that provides both; \tkde{usage based dynamic pricing alongside preserving privacy of smart homes} is required.\\


\begin{table*}[t!]
\begin{center}
 \centering
 \scriptsize
 \captionsetup{labelsep=space}
 \captionsetup{justification=centering}
 \caption{\footnotesize{\textsc{\\ \tkde{A Thorough Analysis of Dynamic Billing and Private Smart Metering Mechanism In Energy Systems.}}}}
  \label{tab:view}
  \begin{tabular}{|P{1cm}|P{0.4cm}|P{0.95cm}|P{2.2cm}|P{2.5cm}| P{1.25cm}|P{2.2cm}|P{1.7cm}|P{1cm}|P{0.8cm}|}
  	\hline
\rule{0pt}{2ex}
\bfseries \centering \tkde{Major Category} & \bfseries \centering Ref No. & \bfseries \centering Focus of Article & \bfseries \centering Mechanism Type & \bfseries \centering \tkde{Functioning of Mechanism} & \bfseries \centering Privacy Type & \bfseries \centering Metrics Enhanced & \centering \bfseries Attacks Tackled   & \centering \bfseries Simulation Platform & \bfseries \modify{Compl-\newline exity} \\
\hline

\multirow{3}{*}{\parbox{2cm}{\centering \textbf{}}}

\rule{0pt}{2ex}
& ~\cite{litref01} &  Dynamic billing & UDP: Usage based dynamic pricing & \tkde{Price control \& aggregation via distributed community gateway }\& price control &  Homomo-\newline rphic encryption & \tabitem Pricing Model &  \tabitem Privacy violation attack & $-$ & $O(n/2)$\\
\cline{2-10}

\rule{0pt}{2ex}
\bfseries \centering Dynamic Pricing & ~\cite{tkdenew14} &  \mubtkde{Price optimization for smart communities}  & Proposed a day-ahead real-time hourly pricing strategy & Used the notion of past distribution to calculate day-ahead and real-time prices & \centering $-$ & \tabitem Energy Price \newline \tabitem Power to Average Ratio & \centering $-$ & \centering $-$ & $-$ \\
\cline{2-10}

\rule{0pt}{2ex}
& ~\cite{tkdenew05} &  \mubtkde{Data aggregation \& dynamic billing}  & \mubtkde{Developed a private aggregation \& billing model for V2G network} & Factoring \& homomorphic encryption based privacy and dynamic billing & Homomo-\newline rphic encryption & \tabitem Computational cost & \tabitem Impersonation attack & \tabitem PBC \newline \tabitem MIRACL  & $-$  \\
\cline{2-10}

\rule{0pt}{2ex}
& ~\cite{tkdenew15} &  \mubtkde{Dynamic pricing for energy trading}  & \mubtkde{Developed a dynamic pricing model to incentivize energy suppliers} & Contract theory based pricing to incentivize users cooperating during peak time & \centering $-$ & \tabitem Energy cost \& demand & \centering $-$ & \centering $-$  & $-$  \\
\cline{2-10}

\rule{0pt}{2ex}
& \cite{newlit01} &  \modify{Dynamic pricing under thresholding policies} & \modify{Developed two optimal dynamic pricing mechanisms } & \modify{Greedy} \& \modify{Sliding-Window heuristic for price developed according to power demand} &  $-$ & \tabitem \modify{Approximation Ratio} \newline \tabitem \modify{Execution Time} &  $-$ & \modify{Java CPLEX} & \textit{\modify{Multiple}}  \\
\cline{2-10}

\rule{0pt}{2ex}
& \cite{tkdenew01} &  \tkderev{Dynamic Energy Prices} & \tkderev{Multi-Objective Optimization for sellers and demanders} & \tkderev{Stake Holders preference based dynamic pricing and demand response model for energy systems.} &  $-$ & \tabitem \tkderev{Energy Price} \newline \tabitem \tkderev{Demand Side Cost} &  $-$ & \tkderev{$-$} & \textit{\tkderev{$-$}}  \\
\cline{2-10}

\hline

\multirow{3}{*}{\parbox{2cm}{\centering \textbf{}}}

\rule{0pt}{2ex}
& ~\cite{tkdenew13} &  \mubtkde{Private smart metering} & Protecting smart metering data via correlated noise & Integrated notion of correlated noise via deep learning for smart meters & Differential privacy & \tabitem MSE \newline \tabitem F-test & \centering $-$ & \centering $-$ & $O(n)$   \\
\cline{2-10}

\rule{0pt}{2ex}
\bfseries \centering Private Grid Reporting & ~\cite{tkdenew07} &  \mubtkde{Private smart metering} & Differentially private noise cancellation for private reporting in smart grid & Multi-master smart meter based noise splitting and cancellation strategy for usage reporting & Differential privacy & \tabitem MAE \newline \tabitem Data leakage & \tabitem Collusion attack \newline \tabitem Correlation attack & Python & $O(n)$  \\
\cline{2-10}

\rule{0pt}{2ex}
\rule{0pt}{2ex}
& ~\cite{litref06} &   \tkde{Protecting peak data and RER data} & \tkde{DPLM: Differentially Private usage monitoring with RER} & \tkde{Integrated differential privacy with intermittent RERs to preserve real-time usage reporting privacy} &  Differential privacy & \tabitem Load usage profiles & \tabitem Eavesdropping attacks & Python & $O(n)$  \\
\cline{2-10}

\rule{0pt}{2ex}
& \cite{tkdenew02} &  \tkderev{Private AMI communication} & \tkderev{Homomorphic encryption based computational friendly privacy preserving} & \tkderev{Multi-category aggregation supported fault-tolerant protocol for smart meters} &  \tkderev{Homomorp- hic Encryption} & \tabitem \tkderev{Computational Cost} &  \tabitem \tkderev{Plaintext attack} & \tkderev{$-$} & \textit{\tkderev{Multiple}}  \\
\cline{2-10}

\hline

\centering \textbf {Private Dynamic Billing} & This Work & \tkde{Incetivized private dynamic Billing} mechanism &  DRDP: Differenially private Private Billing with Usage based Pricing & \tkde{Differential privacy protection for smart homes along with incentivizing cooperative users via dynamic pricing} & Differential privacy & \tabitem Network-wide Privacy \newline \tabitem Usage based billing \newline \tabitem \mubtkde{Benefiting Cooperative Users} & \tabitem \mubtkde{Filtering attack} \newline \tabitem \mubtkde{Data Linking attack} & Python & $O(n)$  \\
\cline{2-10}

\hline

 \end{tabular}
  \end{center}
\end{table*}


\mubtkde{In this paper, we first develop a dynamic pricing} strategy that facilitates \tkderev{the cooperative users and only charges the users who are responsible to cause that peak factor. In order to do so, we work over carrying out private data analysis that effectively tracks whether the user is responsible for peak factor or not.} Furthermore, to ensure privacy in the proposed strategy, we use the notion of differential privacy that adds independent and identically distributed (i.i.d) noise in the real-time metering values to preserve the privacy. The noise is added in such a manner that the data is still useful for billing, DSM, or load-forecasting. \mubtkde{In this regard, we propose a noise adjustment} method to maximize utility alongside preserving privacy. However, it is ensured that NILM techniques will not be able to analyse the exact usage/appliance pattern due to added noise. Collectively, we propose \textbf{D}emand \textbf{R}esponse enhancing \textbf{D}ifferential \textbf{P}ricing (DRDP) mechanism that is responsible for both; private data reporting and usage based dynamic pricing. Experimental evaluation of our proposed DRDP mechanism shows that our mechanism incentivizes cooperative users by only charging the peak price to the users responsible for causing peak value along with providing the benefit of private reporting to smart grid utility. 


\noindent The \mubtkde{remainder of our paper is organized as follows;} \tkde{section 2 provides discussion about previous literature and other state-of-the-art works}, section 3 provides detailed discussion about system model, adversary model, and problem formulation, section 4 provides comprehensive discussion about proposed DRDP mechanism and its algorithmic foundation, section 5 covers all aspects of performance evaluation of DRDP, \tkde{after that, the article is concluded in section 6} by providing concluding remarks and future directions.

\section{Literature Review}

In current literature, certain works highlight the use of dynamic pricing in usage based scenarios, for example, the most prominent work in this domain has been carried out by Liang~\textit{et al.} in \cite{litref01}. \tkderev{In this work, \mubtkde{authors proposed usage based dynamic} pricing and presented a model which uses a distributed community gateway for aggregation and price control features.} In order to enhance privacy, authors used homomorphic encryption based privacy. The presented results enhances previous pricing models along with overcoming privacy violation attack such as eavesdropping attack. \mubtkde{Similarly, another work in the field of dynamic billing has been carried out by authors in~\cite{tkdenew14}. The major focus of the article is to incentivize smart home community by providing them advantages of dynamic pricing on the basis of previous load distributions. Authors first proposed the usage of past load distributions to determine day-ahead prices and then discussed solving and evaluating the difference of day-ahead prices with real-time hourly prices. Another work in the similar domain of dynamic pricing has been carried out by authors in~\cite{tkdenew05}. Authors proposed a private dynamic billing and data aggregation strategy for vehicle to grid (V2G) networks. A relevant work from the perspective of incentivizing energy suppliers via dynamic pricing from perspective of energy trading has been presented by authors in~\cite{tkdenew15}. Authors proposed a contract-theory based approach for dynamic energy pricing.} \modify{Another work \tkde{discussing dynamic pricing under} thresholding policies have been carried out by authors in~\cite{newlit01}. Authors developed two optimal dynamic pricing mechanisms and made greedy and sliding window heuristics for dynamic pricing. \tkderev{A very interesting work using the concepts of multi-objective optimization in order to enhance the communication and computation cost for advanced metering infrastructure (AMI) has been carried out by authors in~\cite{tkdenew01}. The work aims to provide a joint-pricing model for multiple smart homes in a dynamic pricing environment. For this joint pricing, authors proposed a framework in which they integrated the notion of energy supplied, energy system operator, and consumer .}} \\
The other direction in literature review is the integration of privacy preservation in real-time reporting to protect smart home users’ privacy. \mubtkde{In order to do so, a work from the perspective of addition of correlated noise addition of differential privacy via deep learning has been \tkde{presented by authors in~\cite{tkdenew13}.} The article provides a novel combination of deep learning generative adversarial networks (GANs) with smart metering obfuscation from perspective of correlated noise.} \mubtkde{Another work in this field has been carried out by Khadija~\textit{et al.} that also covers the similar domain of integration of differential privacy with smart meter reporting~\cite{tkdenew07}. Authors proposed an efficient noise splitting and cancellation approach with the help of a master smart meter and aggregator.} A work that discussed integration of differential privacy \tkde{for smart meters with renewable energy resources (RER) for real-time smart metering has been presented} by authors in~\cite{litref06}. \tkderev{Another interesting work focusing over the usage of homomorphic encryption scheme to preserve privacy during \mubtkde{smart metering aggregation is presented in~\cite{tkdenew02}. This work supports} multi-part aggregation via preserving privacy in a manner that even if the collecting body or the gateway turns malicious, it will still be able to provide protection.} A table for detailed comparative analysis of all the mentioned mechanisms have been given in~Table~\ref{tab:view}.\\\
\textit{After \tkderev{carrying out careful analysis of all the previous works, it can be summarized that to the best of our knowledge,}} no work that integrates the notion of differential privacy with cooperative users based real-time dynamic billing have been carried out in the literature. \tkde{Similarly, in the preliminary work of this article~\cite{myref01}, we analyze the aspect of dynamic pricing and differential privacy on real-time smart metering data. \mubtkde{In this extended version,} we further propose a noise balancing mechanism for private billing, which can serve as a step forward in the direction of incentivizing users and enhancing demand response along with providing them strong privacy guarantees via differential privacy.}

\section{\tkde{Providing Differentially Private Dynamic Billing}}

In this section \mubtkde{we demonstrate the motivation, problem formulation, system model, and} adversary model for our proposed DRDP mechanism.
\subsection{\modify{Motivation of DRDP}}
\modify{The motivation for the proposed DRDP mechanism is given below:}
\begin{itemize}
\item \modify{Conventional dynamic pricing mechanisms do not incentivize cooperative users and charge the same price to all users within a specific area. We propose a dynamic billing strategy that only charges the users responsible to cause peak factor.}
\item \modify{Traditional dynamic billing strategies does not incorporate the notion of differential privacy to preserve privacy during dynamic billing. However, in our DRDP strategy, we modified the approach of dynamic billing and integrated differential privacy as a privacy preserving notion.}
\end{itemize}

\subsection{Problem Formulation}
We divide the problem formulation of our proposed DRDP mechanism into two parts: first we discuss the privacy requirements for dynamic billing and then we propose three questions that summarizes the problem formulation of our DRDP model.
\subsubsection{Privacy Requirements for Dynamic Billing Scenarios}
Traditional dynamic billing strategies do not incorporate the phenomenon of preserving privacy of homes because they are more concerned towards providing dynamic billing incentives. However, these approaches can raise serious concerns towards privacy of smart homes. Because nowadays, grid utility collect these real-time values in order to predict future load along with management of demand response, but these real-time values can leak personal information of smart home users. For instance, these values can be fed to \tkde{NILM techniques that can even predict appliance usage of a specific house in a specified time-slot. Therefore, it is important to integrate privacy preservation mechanisms in dynamic billing strategy to preserve privacy.} In order to do so, \tkderev{we integrate the notion of differential privacy with smart grid dynamic billing and propose our DRDP} mechanism in this article. 
\subsubsection{Problem Questions}
\modify{We further divided the problem definition of DRDP mechanism  into three critical points mentioned as follows:}
\begin{itemize}
\item \modify{How to incentivize cooperating users that are not responsible \tkde{to cause peak factor in a particular} time-slot? (cf. Section~\ref{IncentiveLabel})}
\item \modify{How to preserve privacy of smart \tkde{meters users alongside giving them} advantages of dynamic billing? (cf. Section~\ref{PrivacyLabel})}
\item \tkderev{How to quantify the probability and expectation of cooperative smart meters in a smart metering network theoretically? (cf. Section~\ref{coopproof})}
\item \modify{How to integrate the \tkde{notion of differential privacy with usage based dynamic billing to provide smart homes with a billing strategy they can trust without worrying about privacy leakage? (cf. Section~\ref{DPPLabel})}}
\end{itemize}

\begin{figure}[t]        
\centering
\footnotesize
\includegraphics[scale = 1]{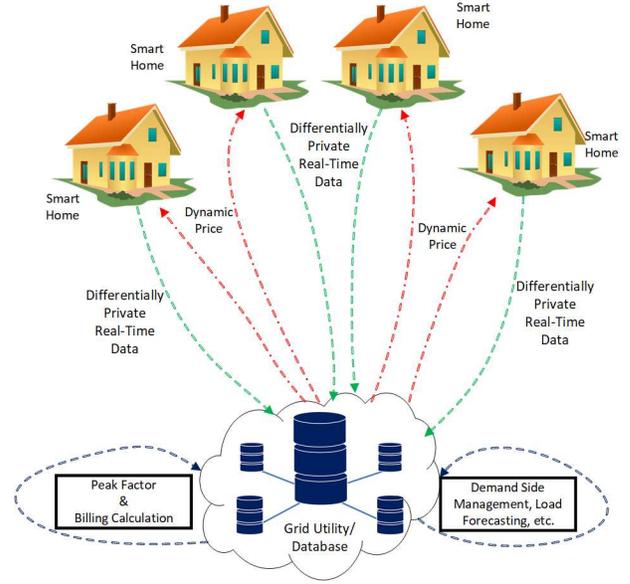}
 \caption{\footnotesize{The proposed system model for DRDP pricing \mubtkde{where each smart meter node in a specified region sends their differentially private} readings to grid utility which further adjusts the noise value via differential privacy adjustment for accurate billing.}}
  \label{fig:dpfig}   
\end{figure}

\begin{table}[t]
\begin{center}
\scriptsize
 \centering
  \captionsetup{labelsep=space}
 \captionsetup{justification=centering}
 \caption{\textsc{\\ \footnotesize{\tkde{Key Notations, Description, and Their Value}}}}
  \label{tab:keynot}
  \color{black}{\begin{tabular}{|P{1.8cm}|P{3.4cm}|P{2cm}|}
  	\hline
  	\textbf{Notation} & \textbf{Description} & \textbf{Value}\\
  	\hline
 	\centering \tkderev{AMI} & \tkderev{Advanced Metering Infrastructure} & -\\
  	\hline
  	\centering \tkderev{DSM} & \tkderev{Demand Side Management} & -\\
  	\hline
  	\centering \tkderev{RER} & \tkderev{Renewable Energy Resources} & -\\
  	\hline
  	\centering \tkderev{DR} & \tkderev{Demand Response} & -\\
  	\hline
  	\centering \tkderev{NILM} & \tkderev{Non-Intrusive Load Monitoring} & -\\
  	\hline
  	\centering $ABS$ & \tkde{Absolute} & -\\
  	\hline
  	\centering $B_R$ & Billing Reading & -\\
  	\hline
  	\centering $F_n$ & \tkde{Function of Noise} & -\\
  	\hline
  	\centering $I_v$ & \tkde{Instantaneous Metering Value} & -\\
  	\hline
  	\centering $D_f$ & Difference Value & -\\
  	\hline
  	\centering $S_c$ & Noise Scale & -\\
  	\hline
  	\centering $N_r$ & Number of Readings & -\\
  	\hline
  	\centering $I_B$ & Instantaneous Bill & -\\
  	\hline
  	\centering $M_n$ & Metering Noise & -\\
  	\hline
  	\centering $G_{SN}$ & Grid Side Noise & -\\
  	\hline  	
  	\centering $t_s$ & Time Slot & -\\
  	\hline
  	\centering $P_v$ & Protected Value & -\\
  	\hline
  	\centering $\mu$ & Mean Value of DP Noise Generation & -\\
  	\hline
  	\centering $P_F$ & \tkde{Peak Factor Value} & 12000Wh\\
  	\hline
  	\centering $P_P$ & \tkde{Price at Peak Time} & ¢25 \\
  	\hline
  	\centering $U_P$ & Unit Price & ¢10\\
  	\hline
  	\centering $N$ & No. of Smart Meters & 10\\
  	\hline	
  	\centering $\Delta f_1$ & Sensitivity at Meter End & 1\\
  	\hline
  	\centering $\Delta f_2$ & Sensitivity at Grid End & 1\\
  	\hline
  	\centering $\mathsf{S_{c1}}$ & Noise Scale Meter End & - \\
  	\hline
  	\centering $\mathsf{S_{c1}}$ & Noise Scale Grid End & - \\
  	\hline
  	\centering $\varepsilon_1$ & Epsilon \scriptsize{(Privacy Parameter at Meter End)} & $Multiple$\\
  	\hline
  	\centering $\varepsilon_2$ & Epsilon \scriptsize{(Privacy Parameter at Grid End)} & $Multiple$\\
  	\hline
  	\end{tabular}}
  \end{center}
\end{table}

\subsection{System Model}
\tkderev{The proposed system model of our DRDP mechanism comprises two major entities e.g., smart homes and grid utility. Smart homes are entities which use energy sent} by smart generation plants to carry out daily operations.  \mubtkde{Our proposed DRDP model provides protection to smart homes in a decentralized manner, because each smart meter adds differentially private noise to each reading before sending it to grid utility. Grid utility is the entity responsible to receive protected live updates from smart homes after a specified value of time. The grid utility works over adjustment of noisy readings received from smart meters accordingly in order to calculate bills accurately.} Grid utility is also responsible for storing data from all smart meters in their database for future statistical tasks, such as \tkde{DSM, load forecasting, etc}. \\
A detailed system \tkde{model is given in Fig}.~\ref{fig:dpfig}, \tkderev{where every smart home is linked with grid utility for real-time billing and monitoring  purposes.} Smart homes are equipped with smart meters which record and accumulate their real-time usage as instantaneous values ($I_v$). After every 10 minutes, smart meters compute a differentially private noise ($M_n$) from Laplace distribution and \tkde{add the generated noise to ($I_v$) to get the protected} metering value ($P_v$). Afterwards, smart meters report this \tkde{protected metering value ($P_v$) to grid utility for billing and other statistical} operations. Grid utility has two major operations, first is calculation of dynamic billing and other is carrying out statistical analysis. \\
In the first operation, grid utility provides fair dynamic pricing to all smart homes depending upon their usage. Grid utility first works over adjustment of these reported values to find the appropriate billing value. Afterwards, it gathers all real-time values ($P_v$) in an specified area and calculates their sum to determine whether the usage for the specific area is larger than peak value or not, in case if its larger than peak value, then it notifies all smart homes that peak-factor is in place, and warns smart homes to use minimum amount of energy. \tkderev{Moreover, it also keeps track of whether a specific house is responsible for causing peak-factor or not.} In case, if a \tkde{specific smart home is consuming larger than average electricity value, then that smart home will be} charged peak-price. Otherwise, the participating houses are only charged the normal price. \mubtkde{A detailed demonstration about this price calculation is given in the DRDP algorithm (cf. Section~\ref{DPPLabel}).} \\
In the second operation, grid utility carries out all statistical tasks along with managing load for all areas. Grid utility manages collected real-time usage data to formulize load curves for future load. \tkderev{Similarly, it also manages grid power stations and provides required instructions regarding different billing scenarios according to each area.}

\subsection{Adversary Model}
\modify{Adversary in our model can be an intruder that is trying to understand the real-time usage pattern of smart homes by analyzing their reported readings ($P_v$). To demonstrate it further, adversaries are actually interested to find out more about the lifestyle of smart homes users. \tkderev{Adversaries can be of two types: 1) Harmless adversaries, who are just interested to know usage patterns to carry out harmless tasks such as targeted advertising after getting information about any damaged appliance in a smart home. These sorts of adversaries collect information of smart homes and feed this information to NILM models, from where they get information that a particular device/appliance is not functioning up to its 100\% capacity and is likely to get damaged soon. In this way, the advertisers start to show the advertisements of a specific product to the targeted customers. Alongside this, certain other aspects, such as price increase for a specific region, etc. also falls in this category. 2) Harmful adversaries, who can cause serious threats to the lives of smart home users and can analyze the valuations to carry out unethical tasks such as burglary, and theft, etc. These sorts of adversaries could be any malicious intruders or hackers, who try to get into databases of smart grid utilities in order to figure out which household is using a specific amount of energy at a specific time. In this way, they try to get information that whether a house is occupied or is empty in a determined time-slot, so that they can perform malicious acts.}}\\
\modify{We further divide the adversarial attacks in our DRDP mechanism into two categories: 1) \tkde{external attack from adversaries, in which adversarial attacker attacks the link of communication between smart home and smart grid utility in order to} find out detailed usage information about homes in a specific region. 2) Internal adversarial attack, in which some internal grid entity acts as an adversarial body and misuses the collected data from grid utility. Since grid utility databases have a large amount of data from all local regions, they can pose large harm in case they act as adversaries. Furthermore, in this scenario, we assume that the adversary is curious-but-honest, as it will not modify, nor will alter or delete the received smart homes readings.}

\begin{algorithm}[t]\small
\caption{\fin{\mubtkde{Smart Meter part of DRDP Algorithm}}}
\label{algoDP1}
\begin{algorithmic}[1]


\State $\textbf{Input} \gets F_n, I_v, \varepsilon_1, \mu, \Delta f_1$ 
\State $\textbf{Output} \gets P_v$

\item[]//Each smart meter will calculate noise as follows:
\item[] $\textbf{FUNCTION} \rightarrow$ DP_Reporting$(I_v, \varepsilon_1, \mu, \Delta f_1)$

\State \textbf{Read} Smart Meter Reading after Specified Interval ($I_v$)
\State \textbf{Initialize} Mean ($\mu$), epsilon ($\varepsilon_1$), sensitivity $\Delta f_1$
\State \textbf{Calculate} Scale $\mathsf{S_{c1}} = \frac{\Delta f_1}{\varepsilon_1}$
\State \textbf{Calculate} Noise = $Lap(I_{v_i}, \mu, \mathsf{S_{c1}})$
\State \textbf{Set} Meter Noise = $M_n = ABS[Lap(I_{v_i}, \mu, \mathsf{S_{c1}})]$
\State \textbf{Set} Protected Value = $P_v = I_{V_i} + M_n$

\State \textbf{return} $P_v$

\item[]//\mubtkde{Protected reading is then sent to grid utility by each smart meter individually.}

\end{algorithmic}
\end{algorithm}


\begin{algorithm}[t]\small
\caption{\fin{\mubtkde{Grid Utility part of DRDP Algorithm}}}
\label{algoDP2}
\begin{algorithmic}[1]

\State $\textbf{Input} \gets P_v, N, P_F, U_p, P_p, I_v, \varepsilon_2, \mu, \Delta f_2$ 
\State $\textbf{Output} \gets B_R, I_B, D_f$
\item[] \;
\item[]//Grid utility will balance noise as follows:

\item[] $\textbf{FUNCTION} \rightarrow$ DPNoiseAdjustment$(N, P_v, \varepsilon_2, \mu, \Delta f_2)$

\For {\texttt{(each \textbf{i} in \textbf{N})}}
\State \textbf{Initialize} Mean ($\mu$), epsilon ($\varepsilon_2$), sensitivity $\Delta f_2$
\State \textbf{Initialize} Protected Value ($P_v$)
\State \textbf{Calculate} Scale $\mathsf{S_{c2}} = \frac{\Delta f_2}{\varepsilon_2}$
\State \textbf{Calculate} Noise = $Lap(P_{v_i}, \mu, \mathsf{S_{c2}})$

\State \textbf{Set} Grid Side Noise = $G_{SN} = ABS[Lap(P_{v_i}, \mu, \mathsf{S_{c2}})]$
\State \textbf{Set} Bill Reading = $B_{R_i} = P_{v_i} - G_{SN}$

\EndFor
\State \textbf{return} $B_{R}$

\item[]//$B_{R}$ is then used to carry out dynamic billing:
\item[] \;

\item[]//Grid utility will carry out dynamic billing as follows:

\item[] $\textbf{FUNCTION} \rightarrow$ DynamicBilling$(N, P_F, B_R, U_p, P_p)$

\For{\texttt{(each \textbf{i} in \textbf{N})}}
        \State \textbf{Set} Sum = $\sum B_{R_i}$
\EndFor

\If {Sum $\geq$ $P_F$}
	\State \textbf{Set} $Avg = P_F/N$

	\For{\texttt{(each \textbf{j} in \textbf{N})}}
        \If {$I_{V_j} \geq  Avg$}
        	\State~$I_{B_j} = B_{R_j} * P_P$
        	\State~\modify{$D_f = B_{R_j} - Avg$}
        \Else
        	\State~$I_{B_j} = B_{R_j} * U_P$
        	\State~\modify{$D_f = Avg - B_{R_j}$}
        \EndIf 
\EndFor
\Else
\For{\texttt{(each \textbf{K} in \textbf{N})}}
\State	$I_{B_k} = I_{B_k} * U_p$
\EndFor
\EndIf
\State \textbf{return} \modify{$I_{B}, D_f$}

\item[]//$I_{B}$ is the price charged to specific user.

\end{algorithmic}
\end{algorithm}


\section{DRDP Mechanism and Its CORE Functionalities}
\subsection{Preliminaries of DRDP}

\subsubsection{Differential Privacy}
The notion of noise addition based privacy preservation also known as differential privacy was first introduced by Cynthia Dwork in 2006 as a means to protect database privacy~\cite{intref06, addref01}. Differential privacy works on the concept of addition of i.i.d noise to obstruct malicious adversaries from recovering private data from sensitive datasets~\cite{tkderef04}. The notion was first used in statistical \mubtkde{databases, but later it was identified that it also provides fruitful} results when it is used on real-time data~\cite{hassanref01}. In this article we use i.i.d noise generated from \tkde{Laplace differential privacy mechanism to preserve  smart metering real-time data privacy}. The formal definitions of differential privacy are as follows:\\
\textbf{Definition 1 (Adjacent Datasets)}\\
\tkde{In a given database $D^n$ consisting of n-dimensions, a query function $Q$ will provide $\varepsilon$-differential privacy $P_d$ if $\forall I_1, I_2 \in D^n$ vary by only a single element and all elements of $R \in range(Q)$~\cite{addref02}.} \mubtkde{Where $R$ is the output value, $D$ is designated database, and $Q$ is the requested function of query that satisfies $\varepsilon$-differential privacy~\cite{intref06}.}
\begin{equation}
\simi{P_{d}[Q(I_1) \in R] \leq e^\varepsilon \times P_{d}[Q(I_2) \in R]}
\label{eqn:eqn1}
\end{equation} 
In the above, $range(Q)$ is the possible range for output value of function $Q$. Correspondingly, the term $\varepsilon$ is the privacy parameter used to determine the amount of noise which is directly linked with the privacy level~\cite{tkderef06, tkderef07}. From perspective of real-time data obfuscation of smart grid, we use the concept of point-wise differential privacy, which was first introduced by Eibl~\textit{et al.} in~\cite{intref08}. \\
\textbf{Definition 2 (Point-wise Sensitivity)}\\
\mubtkde{In traditional differential privacy, sensitivity is usually the smallest difference between two neighboring datasets, however, in real-time scenarios, each individual value is dealt separately. Every real-time value can be counted as an independent entity, and this value can be obfuscated individually on the basis of its current attributes without linking it with its neighbouring value. The formal equation for traditional differential privacy sensitivity can be equated for point-wise sensitivity as follows~\cite{intref08}:}
\begin{equation}
\Delta_{PW} (f) = \max_{t_s,i_1,i_2} |f_{t_S}(i_1)- f_{t_s}(i_2)| = \max_{i,t_s} |X_{i,{t_s}}|
\label{eqn:eqn2}
\end{equation}

\mubtkde{In the above equation, $\Delta_{PW} (f)$ demonstrate the formulation of point-wise sensitivity. First, from traditional neighbouring datasets $f_{t_S}$ perspective, and then from point-wise sensitivity perspective. Similarly, $X_i$ is the value which will be obfuscated respect to differential privacy model.} In our DRDP mechanism, data obfuscation is carried out using the concept of point-wise obfuscation mentioned in Eq.~\ref{eqn:eqn2}. Furthermore, the sensitivity parameter ($\varepsilon$) controls the level of noise for any particular smart meter in a specific time slot ($t_s$). The value of $\varepsilon$ can be varied according to the need, however, this value cannot be taken negative. For interested audience, a more \tkde{detailed discussion regarding differential privacy can be found in }~\cite{intref07}.

\subsubsection{Demand Response \& Dynamic Billing}
\modify{DSM can formally be defined as a method to alter smart home usage profiles in order to match them with the energy supplies~\cite{newcore01}. Similarly, DSM techniques are also being used to reduce operational cost, overcoming black outs, and to reduce emissions of CO$_2$~\cite{newcore02}. Among all DSM mechanisms, DR management is considered to be the most prominent one to maintain a balance between load and supply curve. DR programs are designed and deployed in modern smart grids to enhance participation of smart homes in load balancing. Many types of DR mechanisms have been discussed in literature such as control based, offer based, and decision variable based~\cite{intref01}. Among these mechanisms, offer based mechanisms get a significant amount of attention because they directly incentive users and users can directly see their participation~\cite{tkderef09}.}\\
\modify{In offer based DR models, motivation is developed among smart homes to use minimal amount of energy in the given time slot so that grid utility can balance the load curve and can predict the load in the most proficient manner~\cite{newcore03}. In this article, we use a subcategory of offer based DR mechanism in which we provide incentives to cooperative users on the basis of the factor that determines whether they are contributing in causing peak factor or not.}

\subsection{Functioning of DRDP Mechanism}
\label{DRDPSec}
\subsubsection{DRDP Algorithm} \label{DPPLabel}
\mubtkde{The proposed DRDP algorithm can further be split into two parts, one part is executed at each smart meter individually, while the second part is executed at grid utility end. In this section, we discuss these parts from a technical perspective.}

\paragraph{DRDP Private Reporting}  \label{PrivacyLabel}
\mubtkde{In order to protect the instantaneous values ($I_v$) of smart meters, we use the phenomenon of differentially private noise addition using Laplace differential privacy mechanism. The pseudo-code for noise addition is given in the Algorithm~\ref{algoDP1}. The noise via Laplace is calculated as follows~\cite{intref06}:}

\begin{equation}
\label{lapeq1}
Lap(I_v, \mu, \mathsf{S_{c1}}) = f(I_v, \mu, \mathsf{S_{c1}}) = \frac{1}{2\mathsf{S_{c1}}} e^{\frac{|I_v - \mu|}{\mathsf{S_{c1}}}}
\end{equation}
\mubtkde{Similarly, the above equation can further be broken down for detailed understanding by substituting the value of $(\mathsf{S_{c1}} = \frac{\Delta f_1}{\varepsilon_1})$, the new equation will be~\cite{litref06}}:
\begin{equation}
\label{probeqn}
\resizebox{0.32\hsize}{!}{$f\left(I_v; \mu,\frac{\Delta f_1}{\varepsilon_1}\right) = $}
\resizebox{0.1\hsize}{!}{$\frac{1}{2\frac{\Delta f_1}{\varepsilon_1}}$}.
e^{\resizebox{0.22\hsize}{!}{$\Bigg(-\frac{|I_v-\mu|}{\frac{\Delta f_1}{\varepsilon_1}}\Bigg)$}}
\end{equation}
\tkderev{The calculated noise is then added into the instantaneous value via each smart meter in order to produce a noisy output as follows~\cite{tkdenew09}:}

\begin{equation}
\sum\limits_{i=1}^N \left( P_{v_i} =   I_{v_i} + ABS[Lap(I_{v_i}, \mu, \mathsf{S_{c1}})]\right)
\end{equation}

\tkde{Finally, this protected noisy value is then sent to the smart grid utility} for billing, storage, and future statistical evaluation. \tkderev{Grid utility first works over adjustment of noisy values for billing calculation and then carries out various statistical analysis over these readings such as carrying out load forecasting, etc.} It is important to highlight \mubtkde{that the protected noisy instantaneous values does not have any significant effect on billing or load forecasting as far as the $\varepsilon_1$ value is maintained accordingly because the proposed noise generation model uses a Laplace distribution, which over the period of time ensures that a uniform amount of noise is being generated. Thus, in long-term perspective the error in the billing is minimal. This aspect is thoroughly demonstrated with the help of simulation experiments given in Section~\ref{PerfSect}.}

\paragraph{Differential Noise Adjustment} 
\mubtkde{The first part of grid utility side of DRDP mechanism is differential noise adjustment,} via this function, grid utility generates a random i.i.d noise at its end and reduces this noise value from the reported reading in order to match the accurate value for billing. Firstly, the noise is generated at the grid utility end by using a similar Laplace noise mechanism. \mubtkde{Usually the epsilon value is the same as that of smart meter end, but it can be varied if required. The formal distribution used at grid utility end is as follows~\cite{tkderef08}:}

\begin{equation}
\label{lapeq2}
Lap(P_v, \mu, \mathsf{S_{c2}}) = f(P_v, \mu, \mathsf{S_{c2}}) = \frac{1}{2\mathsf{S_{c2}}} e^{\frac{|P_v - \mu|}{\mathsf{S_{c2}}}}
\end{equation}

The generated noise is then \tkde{reduced from protected value to generate the final reading value} for billing and future analysis. The equation for this process is as follows:
\begin{equation}
\sum\limits_{i=1}^N \left( B_{R_i} =   P_{v_i} - ABS[Lap(P_{v_i}, \mu, \mathsf{S_{c2}})]\right)
\end{equation}

It is important to mention that it is not compulsory that the newly generated value ($B_{R}$) will always match the original value ($I_v$), because there is always a possibility that the new noise value could be pretty small or pretty large as compared to the original noise value generated at meters' end. In the majority of adjustments, both the original and new generated values are pretty similar. But there will always remain a sense of ambiguity and uncertainty in output values even after removal of noise. This introduction of ambiguity is the actual requirement of any differential privacy mechanism, that an adversary will not be able to predict with confidence regarding presence or absence of any individual. In our scenario, if an adversary even gets the corrected values ($B_{R}$) from grid utility, even then these values will be of no use to NILM mechanisms. As these NILM mechanisms will not be able to predict with confidence regarding presence or absence of any specific appliance in smart homes because of the noise ambiguity factor. \tkderev{On the other hand, this noise adjustment does not have much effect on billing values and results have shown that a very minimal level of error is found in billing, which can be ignored because of being very small.} \mubtkde{To keep the accuracy in the billing, we used the absolute function, which ensures that the noise is always positive at the time of addition or reduction from a reading value. Similarly, it is important to highlight that since the noise is generated via a uniform distribution, therefore, in long-term, the long-term noise generation and reduction (e.g., lets say for a month or 10 days) further reduces the impact of noise on the bills, and the final billing price is approximately equal to original value. However, this work can also be extended further and some sort of mechanism which can even calculate accurate instantaneous bill can also be developed in future.}

\paragraph{Incentivizing Cooperative Homes by Dynamic Billing} \label{IncentiveLabel}
\tkderev{Conventional dynamic pricing models usually work in either of the two directions. One way that conventional models used to follow is to provide the same unit rate at fixed predetermined peak factor timings, e.g., if the peak factor is going to be in place from 05:00PM to 10:00PM, then all households will be charged the same price. The second type of conventional dynamic billing models use readings from a specific region to determine whether peak factor is in place or not, and \mubtkde{in this way, they determine the price which should be charged to the households of that specific area. Apart from the two major models, it is important to highlight that some works discussed the use of load-scheduling to off-peak hours, but then this does not incorporate the notion of dynamic peak hours~\cite{intref01}.} Thus, the major issue in majority of these mechanisms is that they do not consider whether a specific home is causing the peak factor or not. E.g., there is a possibility that a household is using minimal amounts of energy during peak timings, but they are still being charged the high price per reading because the peak factor is in place. Here our DRDP model comes in, \mubtkde{in our proposed DRDP mechanism we only charge homes who are causing the peak factor in the specified geographical region. E.g., a specified number of homes, which falls in the near proximity of grid utility are linked with the specific grid station, which are used to determine the occurrence of peak factor. Thus, if a specific house is not responsible to cause peak factor in that specified region, then he will not be charged peak price.}} Due to this, cooperative users will have a motivation to take part in DSM programs, which eventually will have a beneficial effect on load curve. The second function of our proposed DRDP algorithm (Algorithm. ~\ref{algoDP2}) determines and calculates dynamic bill for each smart home. Firstly, adjusted reading values ($B_{R}$) of all smart homes are collected and the sum of all these values is computed via grid utility $(sum =  \sum_{1}^{N} B_{R})$. Afterwards, grid utility derives the peak factor value ($P_F$) for that specific time-slot according to the load curve given by grid utility. Once $P_F$ is determined, utility compares the summation value ($sum_{B_{R}}$) with peak value ($P_F$) to determine that whether the instantaneous sum of all smart homes exceeds peak or not ($sum_{B_{R}} \geq P_F$). In case \tkde{if  instantaneous sum value is larger than selected peak factor, then the smart meter homes are given a notification that peak factor} is in place and energy usage is being charged according to the peak prices. \tkderev{Along with the peak factor comparison, grid utility also calculates instantaneous average value according to the peak factor and number of smart homes via ($avg = \frac{P_F}{n}$).} The instantaneous value of each smart home $B_{R}$ is then compared with this calculated average and in case if the smart home is using energy higher than average, then they are charged for peak prices, for example if $N*$ are the smart homes using energy larger than average, then the billing will be as follows~\cite{tkdenew05}:
\begin{equation}
\label{peakeq2}
\sum_{i = 1}^{N*} I_{B_i}  = \sum_{i = 1}^{N*}(B_{R_i} \times P_{FP})
\end{equation}
Contrary to this, if some smart home is participating in DSM and is using less energy, then they are charged off-peak price ($U_{OP}$) as follows~\cite{litref01}:
\begin{equation}
\label{peakeq1}
\sum_{i = 1}^{N^p} I_{B_i}  = \sum_{i = 1}^{N^p}(B_{R_i} \times U_{OP})
\end{equation}

\modify{We further add the phenomenon of communicating smart homes regarding the energy difference with respect to average value. For example, if a meter is only using 50W more than peak value, which they can reduce, or a smart home is just 10W short from reaching peak value and they do not want to get into peak zone. In order to do so, we calculate the difference between their instantaneous reading ($B_{R}$) and the average ($Avg$) via ($D_f = B_{R} – Avg$) for peak users and ($D_f = Avg – B_{R}$) for non-peak users. \tkderev{This calculated difference value will then be transmitted to the respective smart home to notify them about their usage.}}

\subsubsection{Systematic \& Theoretical Analysis}

\paragraph{Differential Privacy Analysis}
In order to demonstrate that our proposed noise addition mechanism follows differential privacy guarantee, \mubtkde{we carry out theoretical evaluation.} 
The detailed evaluation is given as follows:
\begin{theorem}{\textit{Differentially private metering reporting function of our proposed DRDP mechanism satisfies $\varepsilon_1$-differential privacy guarantee.}}\\
\label{difPriv01}
\hspace{10mm}\textit{\textbf{Proof:}} Let us consider $M_{n}$ \& $M_{n}^\prime$ $\in N^{|X|}$ in a way that $|| M_{n} - M_{n}^\prime||_1 \leq 1$. The arbitrary string length up to $`i’$ for $M_{n}$ \& $ M_{n}^\prime$ will be $M = \{N_1, N_2, . . , N_i\}$. Thus, given that both $M_{n}$ \& $ M_{n}^\prime$ can further be linked with Laplace distribution via probability density function as $p_{M_{n}} \& p_{M_{n}^\prime}$ respectively. These two probability functions can be compared at given arbitrary string (according to Laplace theorem in~\cite{algorithmbook}) as follows:
\begin{align}
\noindent \frac{p_{M_{n}} \left[M = \{N_1, N_2, . . , N_i\}\right]}
{p_{M_{n}^\prime}\left[M = \{ N_1, N_2, . . , N_i\}\right]}  = ~~~~~~~~~~~~~~~~~~~ \nonumber
\end{align}

\begin{align}
& ~~~~~~~~~~~~~~~~~~~~~\prod_{j=1}^{k} 
\frac{\exp\left(- \frac{\varepsilon_1 |F_n(M_{n})_j - N_{j}|}{\Delta f_1}\right)}
{\exp\left(- \frac{\varepsilon_1 |F_n(M_{n}^\prime)_j - N_{j}|}{\Delta f_1}\right)}\\
& = \prod_{j=1}^{k} 
\exp\left(\frac{\varepsilon_1 ( |F_n(M_n^\prime)_j - N_j| - |F_n(M_n)_j - N_j|)}{\Delta f_1}\right)\\
&\leq \prod_{j=1}^{k} 
\exp\left(\frac{\varepsilon_1 ( |F_n(M_n)_j - |F_n(M_n^\prime)_j |)}{\Delta f_1}\right)\\
&= \exp\left(\frac{\varepsilon_1 ( ||F_n(M_n) - |F_n(M_n^\prime)||)}{\Delta f_1}\right)\\
&\leq \exp (\varepsilon_1) 
\end{align}
\end{theorem}

Thus, the above statements prove that differentially private reporting of our DRDP satisfies $\varepsilon_1$–differential privacy. Since, in real-time reporting, we are taking noise values to accumulate in $I_v$, so, the given differential privacy function following positive side of noise symmetry.

\begin{theorem}{\textit{Differential noise adjustment function of our proposed DRDP mechanism satisfies $\varepsilon_2$-differential privacy guarantee.}}\\
\label{difPriv02}
\hspace{10mm}\textit{\textbf{Proof:}} Let us consider $G_{SN}$ \& $G_{SN}^\prime$ $\in N^{|X|}$ in a way that $|| G_{SN} - G_{SN}^\prime||_1 \leq 1$. The arbitrary string length up to $`i’$ for $G_{SN}$ \& $ G_{SN}^\prime$ will be $G_{S} = \{G_1, G_2, . . , G_i\}$. Thus, given that both $G_{SN}$ \& $ G_{SN}^\prime$ can further be linked with Laplace distribution via probability density function as $p_{G_{SN}} \& p_{G_{SN}^\prime}$ respectively. These two probability functions can be compared at given arbitrary string (according to Laplace theorem in~\cite{algorithmbook}) as follows:


\begin{align}
\noindent \frac{p_{G_{SN}} \left[G_S = \{G_1, G_2, . . , G_i\}\right]}
{p_{G_{SN}^\prime}\left[G_S = \{G_1, G_2, . . , G_i\}\right]}  = ~~~~~~~~~~~~~~~~~~~ \nonumber
\end{align}

\begin{align}
& ~~~~~~~~~~~~~~~~~~~~~\prod_{j=1}^{k} 
\frac{\exp\left(- \frac{\varepsilon_2 |F_n(G_{SN})_j - N_{j}|}{\Delta f_1}\right)}
{\exp\left(- \frac{\varepsilon_2 |F_n(G_{SN}^\prime)_j - N_{j}|}{\Delta f_1}\right)}\\
& = \prod_{j=1}^{k} 
\exp\left(\frac{\varepsilon_2 ( |F_n(G_{SN}^\prime)_j - N_j| - |F_n(G_{SN})_j - N_j|)}{\Delta f_1}\right)\\
&\leq \prod_{j=1}^{k} 
\exp\left(\frac{\varepsilon_2 ( |F_n(G_{SN})_j - |F_n(G_{SN}^\prime)_j |)}{\Delta f_1}\right)\\
&= \exp\left(\frac{\varepsilon_2 ( ||F_n(G_{SN}) - |F_n(G_{SN}^\prime)||)}{\Delta f_1}\right)\\
&\leq \exp (\varepsilon_2) 
\end{align}
\end{theorem}

Thus, the above statements prove that differential noise adjustment function of our DRDP satisfies $\varepsilon_2$-differential privacy. Since, in noise adjustment, we are taking removing noise values in order to match the correct values of $I_v$ as much as possible. So, the given noise adjusting differential privacy function following negative side of noise symmetry.\\
Both Theorems ~\ref{difPriv01} \& ~\ref{difPriv02} can be combined to prove that both side of symmetries of Laplace distribution are followed in our DRDP model.

\begin{lemma}{\mubtkde{Let us consider $Z_1(q)$ and $Z_2(q)$ be two algorithms that are differentially private having their respective privacy budget values $\varepsilon_1$ and $\varepsilon_2$ respectively. Then, $Z(q) = (Z_1(q),Z_2(q))$ satisfies ($\varepsilon_1 + \varepsilon_2$)-differential privacy with respect to the composition theorem demonstrated in~\cite{algorithmbook}. }}
\label{lemmalabel01}
\end{lemma}

\begin{theorem}{\textit{\mubtkde{Our proposed \textbf{D}emand \textbf{R}esponse enhancing \textbf{D}ifferential \textbf{P}ricing (DRDP) mechanism satisfies $\varepsilon$-differential privacy guarantee.}}}\\

\hspace{10mm}\textit{\textbf{Proof:}} In our proposed DRDP algorithm, Laplace distribution of differential privacy is applied in a sequential step-wise manner via $\varepsilon_1$ \& $\varepsilon_2$ privacy budgets. Thus, by following composition theorem of differential privacy according to Lemma 1, if we perform sequential perturbation on same smart metering data by using $\varepsilon_1$ \& $\varepsilon_2$, then, they can be accumulated via summation to prove differential privacy guarantee (e.g., ${\sum\limits_j} \varepsilon_j - dp $). \mubtkde{Therefore, our proposed differential noise addition (using $\varepsilon_1$) and differential noise adjustment (using $\varepsilon_2$) of DRDP can be written as ($\varepsilon_1 + \varepsilon_2$)-differential privacy to prove differential privacy guarantee via composition theorem. Thus, both privacy parameters can be generalized as ($\varepsilon$) in order to state that our proposed DRDP mechanism satisfies $\varepsilon$-differential privacy guarantee.}
\end{theorem}

\paragraph{\modify{Cooperative State Analysis}} \label{coopproof}
Considering the system model and functioning given in previous sections, it can be visualized that at the time of peak-factor in place, two types of behaviors of smart meter nodes can be seen. Either they are in cooperative state (e.g., using less than average) or in non-cooperative state (using more than average). \tkderev{Based on these conditions, we devise two states of the system named as cooperative and non-cooperative.} To clear it further, we developed the notion that when at least half of the smart meters will be in a cooperative state and using less than average value, then the complete network will be in a cooperative state. Contrary to this, if less than half of the smart metering nodes are using more than average, then the system will be in a non-cooperative state. \tkderev{In order to quantify a theoretical relation for this cooperative nature, we carry out a detailed theoretical analysis of smart metering systems being in a cooperative state or not. If the total number of smart meters/homes in a specific area are `N’, and only `q’ nodes are cooperating with the grid utility, then the probability and expectation of smart meters in cooperative state can be determined via probability theory binomial random variable analysis. \tkderev{The determined states via probability and expectation analysis can then be used to determine the future response of the system based upon its current status. In this way, smart grid utilities will be able to determine or choose their prospective strategy accordingly. \mubtkde{For instance, if a specific ratio of smart meters are in a cooperative state, then they can provide some significant incentives to those smart meters. Similarly, if a large number of smart meters are not cooperating in a region, then certain penalty scores can be introduced for such scenarios. This direction can also be explored further in order to develop more advanced demand response strategies.}}}
\vskip 2mm
\begin{theorem}{\textit{ The probability of system being in cooperative state is~\cite{tkdenew10}:}}\newline
\label{theoremprob}
\vspace{-2em}
\begin{dmath}
P_{cs} = \sum_{q = \ceil{\frac{N}{2}}}^{N} {N \choose q} \left(P_{LU}^{(m)}\right)^q \left(P_{HU}^{(m)}\right)^{N-q}
\end{dmath}
In the above equations, $P_{LU}$ and $P_{HU}$ are cooperative and non-cooperative user probabilities respectively which are demonstrated in detail below. \newline
\hspace{10mm}\textit{Proof:} Considering the factor that a smart meter can either be in a cooperative or a non-cooperative sate we determine the state probability vectors as follows~\cite{tkdenew10}:
\[
P_{LU} = \{P_{L(1)}, P_{L(2)}, P_{L(3)}, . . . , P_{L(N)}  \}
\]
\[
P_{HU} = \{P_{H(1)}, P_{H(2)}, P_{H(3)}, . . . , P_{H(N)}  \}
\]

\mubtkde{Consider $S_m$ be the binomial random variable for smart meters in a cooperative state. Thus, P($S_m = q$) will be the probability that $q$ number of nodes in cooperative state during peak-hours, which can be written as follows~\cite{tkdenew10}:} 

\begin{dmath}
\label{eqnvalue01}
P\{S_m = q\} = {N \choose q} \left(P_{LU}^{(m)}\right)^q \left(1 - P_{LU}^{(m)}\right)^{N-q}
\end{dmath}

The system will remain in non-cooperative state unless $\ceil{\frac{N}{2}}$ smart meters enters in cooperative state, so, the probability of being in non-cooperative state can be calculated from Eq.~\ref{eqnvalue01} as follows~\cite{tkdenew10}:

\begin{dmath}
P_{NC} = \sum_{q=0}^{\floor{\frac{N}{2}}} {N \choose q} \left(P_{LU}^{(m)}\right)^q \left(1 - P_{LU}^{(m)}\right)^{N-q}
\end{dmath}

\mubtkde{Complying with the probability condition that ($P_{CS} + P_{NC}  = 1)$~\cite{tkdenew10}, the above equation can be written as~\cite{tkdenew10}:}

\begin{dmath}
\label{eqnvalue02}
P_{CS} = 1 - P_{NC} = 1 -  \sum_{q=0}^{\floor{\frac{N}{2}}} {N \choose q} \left(P_{LU}^{(m)}\right)^q \left(1 - P_{LU}^{(m)}\right)^{N-q}
\end{dmath}
According to the probability vectors of $P_{LU}$ and $P_{HU}$, individual values of each vector can be compared according to the probability condition of summation equal to 1 (e.g., $P_{L(1)} + P_{H(1)} = 1$), which can further generalized for above summation as $P_{LU}^{(m)} + P_{HU}^{(m)} = 1$. So, Eqn.~\ref{eqnvalue02}, will become:

\begin{dmath}
\label{eqnvalue03}
P_{CS} = 1 -  \sum_{q=0}^{\floor{\frac{N}{2}}} {N \choose q} \left(P_{LU}^{(m)}\right)^q \left(P_{HU}^{(m)}\right)^{N-q}
\end{dmath}

The above equation provides the probability of system being in cooperative state, which means that at least $\ceil{\frac{N}{2}}$ nodes are in cooperative state. So, the Eqn.~\ref{eqnvalue03} can be modified to prove the theorem as follows~\cite{tkdenew10}:

\begin{dmath}
\label{finalprobability}
P_{cs} = \sum_{q = \ceil{\frac{N}{2}}}^{N} {N \choose q} \left(P_{LU}^{(m)}\right)^q \left(P_{HU}^{(m)}\right)^{N-q}
\end{dmath}

\end{theorem}

Moreover, Eq.~\ref{finalprobability} can be used to determine the expected value of smart meters, \mubtkde{which is used to determine the expected number of smart meter nodes in cooperative state at different probability values (according to expectation case of random variable~\cite{tkdenew10}).} So, the equation for expectation can be derived from Eq.~\ref{finalprobability} as~\cite{tkdenew10}:

\begin{dmath}
E[P_{CS}] = \sum_{q = \ceil{\frac{N}{2}}}^{N} q. {N \choose q} \left(P_{LU}^{(m)}\right)^q \left(P_{HU}^{(m)}\right)^{N-q}
\end{dmath}

\tkderev{From the above equations, one can determine a probability for smart homes which will be in a cooperative state in a particular time frame, alongside determining the expected value for cooperative smart homes.  }

\paragraph{\modify{Complexity Analysis}}
\modify{\newline The proposed DRDP Algorithm for real-time private reporting and smart dynamic billing provides an efficient solution as it only utilizes minimal required amount of operations for its execution. The theoretical proof for this analysis is as follows: }

\begin{theorem}{\textit{The computational complexity of our proposed DRDP Algorithm has an upper bound of $\mathcal{O}(N)$ because the algorithm iterates a maximum `N' number of times. \mubtkde{Similarly, the lower bound at smart meter side will be $\mathcal{O}(1)$, because it will only take a single step to add noise and report it to grid utility. However, the noise balancing/adjustment and billing functions, which runs at grid utility iterates `N’ number of times, which makes the lower complexity bound to be $\mathcal{O}(N)$ at grid utility end.}}}
\label{theorem01}
\end{theorem}

\paragraph{\mubtkde{Privacy Attacks Analysis}}
\mubtkde{Our proposed DRDP mechanism provides resilience against filtering attack and data linking attack. 
Data linking attack is a type of privacy attack in which an adversary tries to predict private data of a user by observing and linking the given information with other similar information from the same of other similar databases~\cite{tkdenew06}. In smart metering data perspective, data from multiple sources is linked with smart meters in order to combine and arrange information in such a way that private information can be inferred. This attack can be carried out via some insider or an external adversary, e.g., an insider reconstruction attack will be the one, where some insider such as grid utility tries to launch this attack over the reported data. However, we can say it with confidence, that our proposed DRDP model provides an effective resilience to such data linking attack, even if gets launched from grid utility end. This is because of the reason that the noise is added locally from smart metering side, and the noisy value is generated via differential privacy in such a manner ensures that the grid utility or any other intruder will not be able to infer into private information of smart meter users, even if it tries to link it with other similar databases. This is because of the strong privacy guarantee provided by the theoretical basis of differential privacy, especially when the privacy budget $\varepsilon$ is taken into consideration appropriately.~\cite{tkdenew08}}.
\mubtkde{Similarly, from perspective of filtering attack, strong statistical analysis and negative noise generation is usually carried out from adversary side in order to get exact reading of smart meters. However, our proposed DRDP mechanism uses strong notion of differential privacy, which ensures that even strong statistical analysis or negative noise generation will not be quite helpful for adversaries. And adversaries will not be able to re-construct the original values from private reported data~\cite{tkdenew09,tkdenew07}.}

\section{Performance Evaluation of DRDP}\label{PerfSect}
\mubtkde{To evaluate our DRDP mechanism, we took the dataset of~\cite{expresult}, and extracted real-time values of randomly picked 10 smart homes in order to carry out our experimentation of DRDP model. Furthermore, we carry out comparison with usage based dynamic pricing presented in the works as UDP~\cite{litref01} and PADP~\cite{tkdenew05}.} Furthermore, \tkde{to perform experimental evaluation, we use NumPy library} NumPy  from Python 3.0, and performed experiments over smart meter transmitted data having an interval of 10 minutes between each reading~\cite{litref06}.  The simulation parameters used in our experiment are provided in Table.~\ref{tab:keynot}.\\
We further divide the experimental evaluation into three parts, first we analyze DRDP strategy from perspective of differential privacy noise addition and adjustment, afterwards, we analyze the dynamic billing, and finally we evaluate the cooperative smart home analysis. 

\begin{figure}[t]        
\centering

\includegraphics[scale = 0.65]{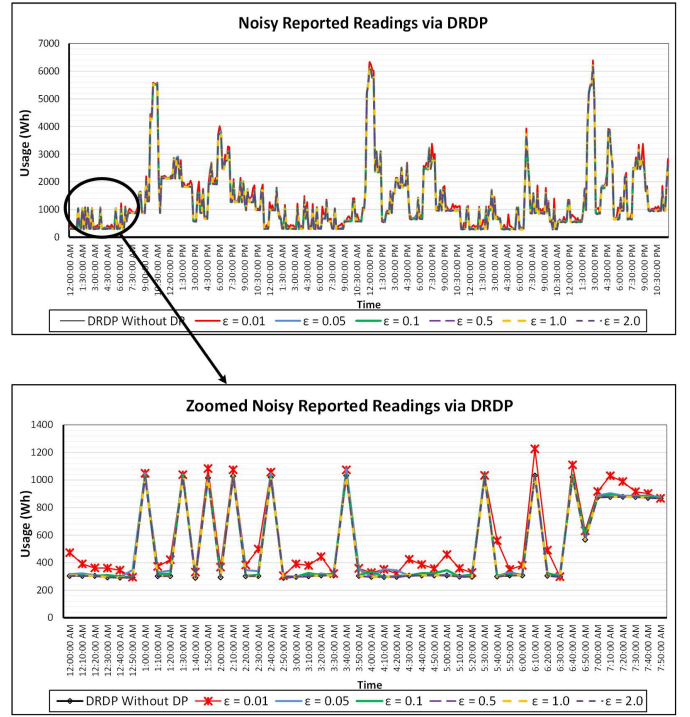}
 \caption{Performance Evaluation of Noisy Reporting Function of DRDP Mechanism. The graph shows absolute private values reported to smart grid utility from smart meter after addition of differentially private noise at different epsilons ($\varepsilon$) values.}
  \label{fig:noisyreportfig}   
\end{figure}

\begin{figure}[t]        
\centering

\includegraphics[scale = 0.63]{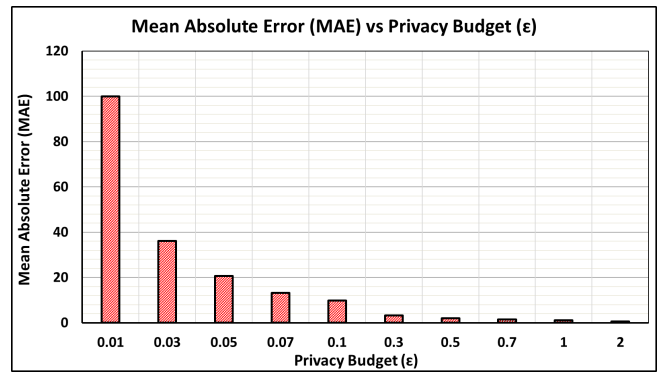}
 \caption{Analysis of Mean Absolute Error (MAE) Added in each Meter Reading with Respect to Privacy Budget ($\varepsilon$). The values of MAE are absolute error values and are not in percentage.}
  \label{fig:nopeaksfig}   
\end{figure}

\begin{figure}[t]        
\centering

\begin{center}

\includegraphics[scale = 0.65]{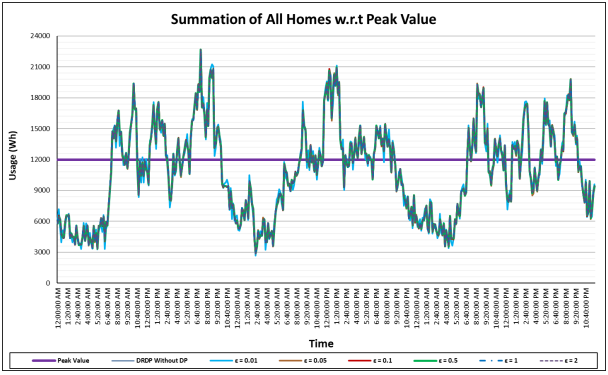}

\caption{Accumulated Sum of All Participating Homes after Noise Adjustment via DRDP.}
\label{fig:SummationBill}
\end{center}
  
\end{figure}

\begin{figure}[t]        
\centering

\begin{center}

\includegraphics[scale = 0.54]{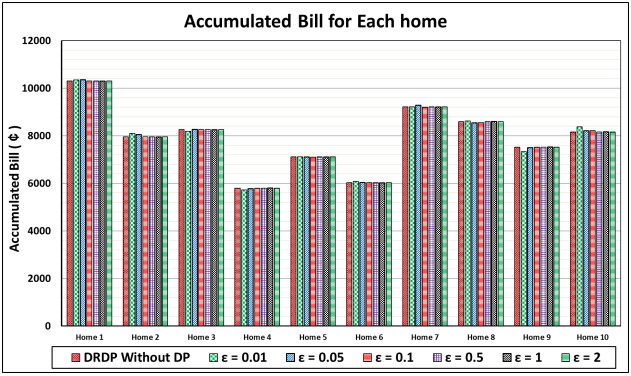}

\caption{Accumulated Billing Sum for 10 Homes Using after using Incentivized Dynamic Billing of DRDP on Adjusted Noise Values Reported at Different Privacy Budgets.}
\label{fig:AccumulatedBillingGraph}
\end{center}
  
\end{figure}

\begin{figure}[t]        
\centering

\begin{center}

\includegraphics[scale = 0.45]{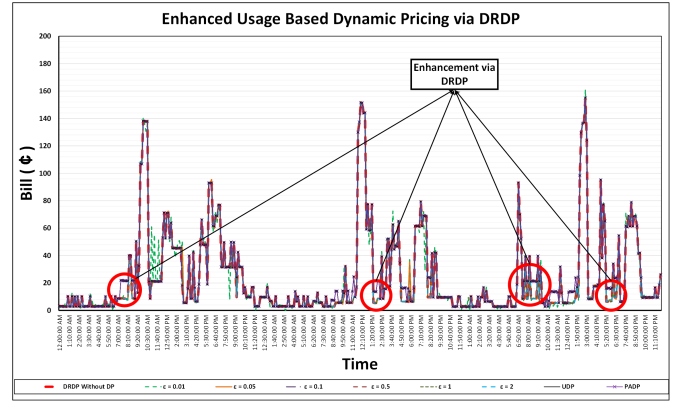}

\caption{\mubtkde{Billing Graphs for a Randomly Picked Smart Home in order to Visualize the incentive given by DRDP as Compared to UDP~\cite{litref01} \& PADP~\cite{tkdenew05}.}}
\label{fig:BillingGraph}
\end{center}
  
\end{figure}

\begin{figure}[t]        
\centering

\begin{center}

\includegraphics[scale = 0.39]{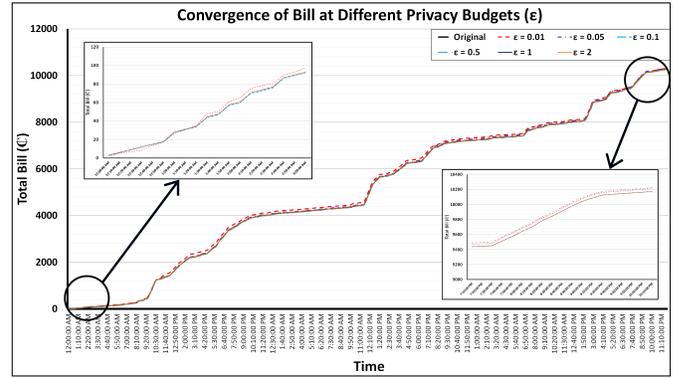}

\caption{\mubtkde{Convergence Graph for a Randomly Picked Smart Home in order to Visualize the Effectiveness of DRDP Billing over a Time Period.}}
\label{fig:Convergence}
\end{center}
  
\end{figure}

\begin{figure}[t]        
\centering

\begin{center}

\includegraphics[scale = 0.58]{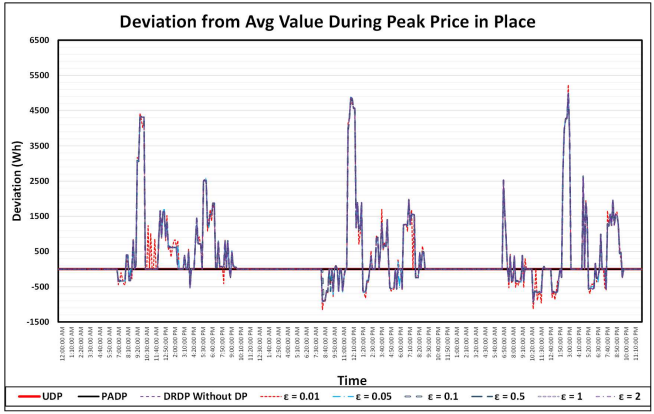}

\caption{\mubtkde{Evaluation of Deviation Notification Function of DRDP from Each Bill reading for a Smart Home.}}
\label{fig:DeviationBill}
\end{center}

\end{figure}

\begin{figure}[t]        
\centering

\includegraphics[scale = 0.73]{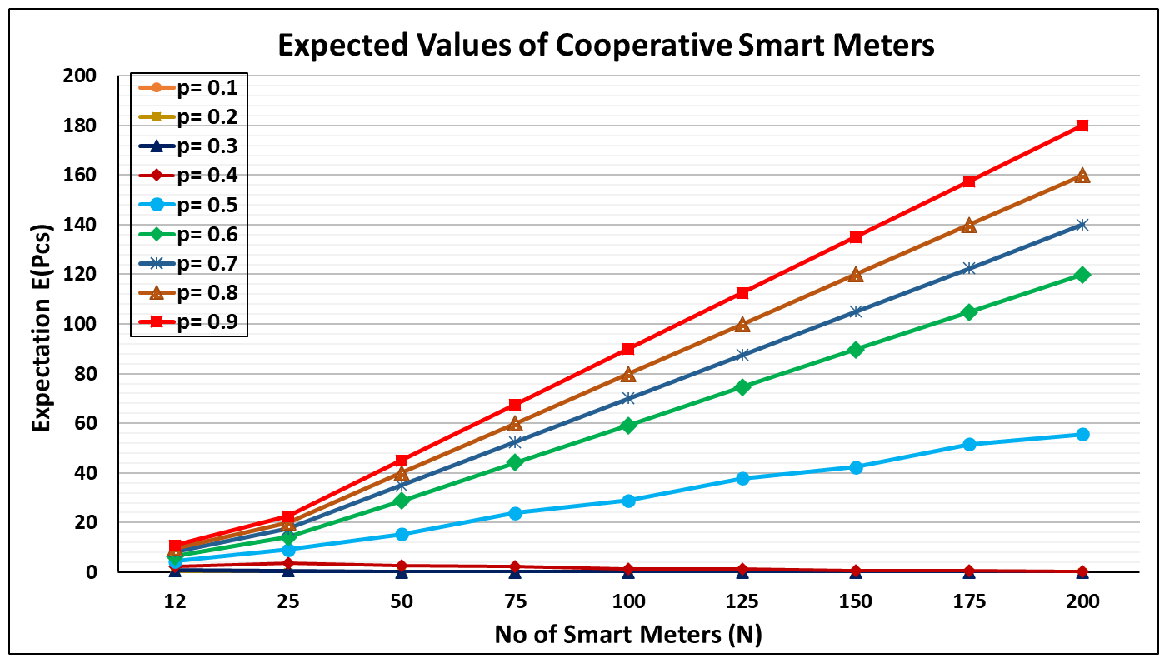}
 \caption{Expectation of Smart Homes for Cooperative State Analysis at Different Probability Values.}
  \label{fig:expectfig}   
\end{figure}

\subsection{Private Grid Reporting and Noise Adjustment}
The graphs presented in Fig.~\ref{fig:noisyreportfig},~\ref{fig:nopeaksfig}, and~\ref{fig:SummationBill} demonstrate the noise reporting and adjustment scenario. Firstly in Fig.~\ref{fig:noisyreportfig}, two graphs are shown, the first graph demonstrates the real-time readings reported from smart meter to grid utility. \tkderev{The graph is built using 3 days of data of smart home readings at different values of privacy parameters. In order to show the effectiveness of our DRDP mechanism, we provide a thorough analysis of real-time reading of smart homes at different $\varepsilon$ values ranging from 0.01 to 2.0 with different intervals. The solid black line shows the reported reading via DRDP without differentially private noise addition, while the other lines demonstrate the noise addition at different privacy budgets. It can be visualized, that at different privacy budgets, the reported value distorts accordingly in order to protect the privacy of smart homes.} The second graph in the figure is the zoomed version of the first graph, which is zoomed in order to visualize the changes due to added noise. From both of the figures, it can be seen that the addition of noise distorts the original values for privacy protection. Especially, when the value of $\varepsilon$ is small, large distortion can be seen at the values, which means more privacy is preserved. Fig.~\ref{fig:nopeaksfig} can be visualized in order to find out the error rate at each $\varepsilon$ value. Mean absolute error (MAE) in our DRDP is calculated by taking the sum of absolute difference between noisy values and original values for a smart home throughout the reported time, and then this accumulated difference is divided by the total readings involved in the experiment (e.g., 3 days in our experiments) ($MAE = \frac{\sum_{n=1}^{N_r}|P_v - I_v|}{N_r}$). \mubtkde{In this way, MAE can be used to determine the error in the reported readings with respect to original reading. From Fig.~\ref{fig:nopeaksfig}, it can be seen that the value of MAE is highest at the time of $\varepsilon$ = 0.01, which means that approximately a difference of 100 is added to reading reported to smart grid utility from smart meter.} Similarly,  this value tends to reduce with the increasing value of $\varepsilon$. It is important to mention that lower values does not mean that privacy is not preserved, as the lower values do also preserve privacy of smart meters from NILM strategies to a greater extent as NILM strategies cannot predict with confidence due to added noise. \\
Moving further to noise adjustment, the graph demonstrating the effect on accumulated meter reading can be seen in Fig.\ref{fig:SummationBill}. The given graph demonstrates the summation of usage of all homes for 3 days after noise adjustment. It can be seen from the graph, that even after summation of values from all 10 homes, very minimal difference can be seen with respect to original value ‘without proposed DRDP’. Which means that the adjusted values are pretty close to the original values, which directly means that the error in the billing value will be very minimal which can be neglected by considering it as a trade-off of preserving privacy. \mubtkde{The summation values are used to determine the occurrence of peak factor in a specified region. E.g., in our experiments, if the accumulated value is more than 12,000Wh, then the peak threshold is reached and users will be notified accordingly.} These adjusted values are then fed to the billing function for bill calculation, which is demonstrated in the next part of performance evaluation.

\subsection{Incentivized Billing Evaluation}
Since, billing is another major aspect of our contribution, so, we demonstrate this functionality by showing experimental results in Fig,~\ref{fig:AccumulatedBillingGraph},~\ref{fig:BillingGraph},~\ref{fig:DeviationBill}. \mubtkde{It is important to mention that multiple tariff plans can be used from perspective of peak and off-peak billing~\cite{tkdenew11}. However, in order to provide our readers a clear understanding, we use standard unit price $U_P$ as ¢10 and peak price $P_P$ as ¢25.} The major concern while calculating billing from noisy values was that it will have huge errors, and this will not be able to match with original values. However, we overcame this concern by proposing a noise adjustment function, and we evaluated its usefulness at different $\varepsilon$ values and showed it in Fig.~\ref{fig:AccumulatedBillingGraph}. The accumulated bill shown in the figure is calculated by accumulating billing values of all smart homes within a timespan of three days. In the given figure, the first bar in red texture shows the proposed dynamic billing strategy but without noise addition. However, the remaining bars show the accumulated reading of a smart home by using noisy values at different privacy budgets. From the results, it can be visualized that there is very minimal difference in bills of all smart homes. Even at $\varepsilon$ = 0.01, when the value of noise is pretty high at the time of reporting, even then the accumulated bill of all smart homes have very low or no variance with respect to the original bill. These results demonstrate the effectiveness of our noise adjustment function, that from surface level one can take the perception that the noisy value might cause bill billing error, but this did not happen. Contrary to this, in the long-run the overall billing difference is negligible. \mubtkde{So, we suggest that our proposed mechanism can be implemented in real-time smart meters \tkde{to protect their privacy along side providing them usage based dynamic billing.}}\\
Furthermore, in Fig.~\ref{fig:BillingGraph}, the separated billing graph for one smart home can be visualized. \tkde{In the given graph, it can be visualized that our proposed DRDP mechanism only charges the user when they are causing the peak factor} and it does not charge when the specific home is not causing peak factor. E.g., from 07:30AM to 09:00AM, it can be seen that the home was not responsible for causing the peak factor, however, the PADP and UDP strategies charged the smart home with peak price. Same result can be seen almost everyday in the time-slot from 07:30AM to 09:00AM, as the home is generally cooperating in these slots because the peak factor has occurred. Therefore, according to DRDP strategy, it is being charged a low price because of its cooperation, however, in UDP mechanism, it is being charged according to the same tariff as that of other smart homes. \mubtkde{From this perspective, the accumulated bill for the specified home via PADP is ¢10,994, while via DRDP without DP is ¢10,301, which is approximately 6.3\%~less comparatively. It is important to mention that this comparison is very specific and oriented towards the picked smart home. There is a possibility that if some smart home is cooperating in majority of peak slots, then the difference in his bills via PADP and DRDP will be pretty high in comparison to a smart home which is not cooperating at all.}\\ 
\mubtkde{Moreover, Fig.~\ref{fig:Convergence} demonstrate effectiveness of our bill calculation algorithm from convergence perspective. It can be visualized that at higher privacy budget values, the bill convergence with respect to original bill begins from the starting values and for lower privacy budget values (such as $\varepsilon$ = 0.01), the error in the bill reduces with the passage of time and is approximately negligible at the end of third day. Thus, our DRDP model provides effective and approximately accurate billing for regimes which does not require instantaneous billing at start.} The next graph (Fig.~\ref{fig:DeviationBill}) shows the output of the deviation function that we added in our enhanced pricing model. This function calculates the difference of a smart meter from the average value and reports the difference to the smart meter user in order for him to take adequate action. E.g., in case if peak factor is in place and a smart home is just using a few watts less than peak value, then it is notified that you are `X’ amount short from reaching average value. This notification is like an initial alert message to smart home users, which afterward try to control its usage a bit further in order to not fall above peak factor.

\subsection{Cooperative State Evaluation}
From the perspective of cooperative state analysis, we provide experiment results in Fig.~\ref{fig:expectfig}. In the given figure, the expected value of the number of smart meters has been shown at different probability values. For example, in the case of 12 smart meters, the expectation is minimum at $p = 0.1$, however, the same values reached approximately the maximum limit at $p=0.9$. The same trend can be visualized for other number of smart meters as well, which can be used to conclude that higher the probability value, higher will be the expected number of smart meters in cooperative state.\\
\noindent \textit{Hence, after careful analysis of the experimental results provided in experimental graphs,  it can easily be determined that  DRDP mechanism efficiently provide smart metering privacy along with providing benefit to cooperative users in dynamic billing scenario.}

\subsection{Discussion}

\mubtkde{Differentially private dynamic billing via noise cancellation is a new direction, and the proposed DRDP model provides an efficient solution to preserve privacy of smart metering users alongside providing them the benefits of cooperative dynamic billing. Alongside this, we believe that it also opens up a window for a large number of future directions and challenges. For instance, attacks such as collusion attack, eavesdropping attack, filtering attack, and data disclosure attacks have also been discussed in the smart metering domain. Currently, these attacks are not analysed, but in future, we are planning to implement and to provide solutions to all of these attacks in the context of the DRDP mechanism. Similarly, since the proposed DRDP model functions over noise cancellation mechanism, thus, it provides an efficient solution to approximately accurate billing for a specified time interval (let us say 1 day or few days). Thus, the error rate reduces evenly and the billing accuracy converges as the number of readings are increased and eventually a negligible error is achieved for billing. However, for instantaneous billing, this approach does not work perfectly and can be extended further to provide the facility of accurate instantaneous bill for a fixed time slot only for a different dynamic billing regime which require instantaneous billing. Alongside this, certain similar works from the perspective of privacy preservation and dynamic pricing (such as~\cite{tkdenew07, tkdenew01, tkdenew14, tkdenew13, tkdenew15}) are present in the literature. In future, we plan to compare these works with our DRDP model in order to propose a more efficient pricing and privacy model for the future smart grids.}

\section{Conclusion}

In this paper, we enhance traditional dynamic billing mechanisms for smart homes by providing an incentivising dynamic pricing mechanism for cooperative users. Furthermore, we provide a differentially private reporting mechanism for smart meters to protect their privacy. Collectively, a \textbf{D}emand \textbf{R}esponse enhancing \textbf{D}ifferential \textbf{P}ricing (DRDP) mechanism has been proposed, which can be incorporated into smart meters and grid utility for efficient demand side management. A detailed theoretical analysis has been carried out for our proposed DRDP mechanism. Similarly, extensive performance evaluation at different privacy parameters have been carried out as well. The provided analysis and performance evaluation show that our proposed DRDP mechanism outperforms traditional and state-of-the-art works in dynamic pricing and private smart metering.\\


\bibliographystyle{IEEEtran}

\begin{thebibliography}{10}
\providecommand{\url}[1]{#1}
\csname url@samestyle\endcsname
\providecommand{\newblock}{\relax}
\providecommand{\bibinfo}[2]{#2}
\providecommand{\BIBentrySTDinterwordspacing}{\spaceskip=0pt\relax}
\providecommand{\BIBentryALTinterwordstretchfactor}{4}
\providecommand{\BIBentryALTinterwordspacing}{\spaceskip=\fontdimen2\font plus
\BIBentryALTinterwordstretchfactor\fontdimen3\font minus
  \fontdimen4\font\relax}
\providecommand{\BIBforeignlanguage}[2]{{%
\expandafter\ifx\csname l@#1\endcsname\relax
\typeout{** WARNING: IEEEtran.bst: No hyphenation pattern has been}%
\typeout{** loaded for the language `#1'. Using the pattern for}%
\typeout{** the default language instead.}%
\else
\language=\csname l@#1\endcsname
\fi
#2}}
\providecommand{\BIBdecl}{\relax}
\BIBdecl

\bibitem{newint01}
P.~Kumar, Y.~Lin, G.~Bai, A.~Paverd, J.~S. Dong, and A.~Martin, ``Smart grid
  metering networks: A survey on security, privacy and open research issues,''
  \emph{IEEE Communications Surveys \& Tutorials}, vol.~21, no.~3, pp.
  2886--2927, 2019.

\bibitem{newint02}
Z.~{Ma}, H.~{Zhong}, Q.~{Xia}, and C.~{Kang}, ``A block-of-use electricity
  retail pricing approach based on the customer load profile,'' \emph{IEEE
  Transactions on Smart Grid}, vol.~11, no.~2, pp. 1500--1509, 2020.

\bibitem{tkdenew14}
A.~Anees, T.~Dillon, S.~Wallis, and Y.-P.~P. Chen, ``Optimization of day-ahead
  and real-time prices for smart home community,'' \emph{International Journal
  of Electrical Power \& Energy Systems}, vol. 124, p. 106403, 2021.

\bibitem{tkderef01}
M.~{Majidi} and K.~{Zare}, ``Integration of smart energy hubs in distribution
  networks under uncertainties and demand response concept,'' \emph{IEEE
  Transactions on Power Systems}, vol.~34, no.~1, pp. 566--574, 2019.

\bibitem{intref01}
J.~S. Vardakas, N.~Zorba, and C.~V. Verikoukis, ``A survey on demand response
  programs in smart grids: Pricing methods and optimization algorithms,''
  \emph{IEEE Communications Surveys \& Tutorials}, vol.~17, no.~1, pp.
  152--178, 2015.

\bibitem{tkdenew03}
D.~Zhang, H.~Zhu, H.~Zhang, H.~H. Goh, H.~Liu, and T.~Wu, ``Multi objective
  optimization for smart integrated energy system considering demand responses
  and dynamic prices,'' \emph{IEEE Transactions on Smart Grid}, pp. 1--1, 2021.

\bibitem{litref01}
X.~Liang, X.~Li, R.~Lu, X.~Lin, and X.~Shen, ``Udp: Usage-based dynamic pricing
  with privacy preservation for smart grid,'' \emph{IEEE Transactions on Smart
  Grid}, vol.~4, no.~1, pp. 141--150, 2013.

\bibitem{tkdenew04}
W.~Tushar, T.~K. Saha, C.~Yuen, T.~Morstyn, Nahid-Al-Masood, H.~V. Poor, and
  R.~Bean, ``Grid influenced peer-to-peer energy trading,'' \emph{IEEE
  Transactions on Smart Grid}, vol.~11, no.~2, pp. 1407--1418, 2020.

\bibitem{tkderef02}
K.~{Chen}, Y.~{Zhang}, Q.~{Wang}, J.~{Hu}, H.~{Fan}, and J.~{He}, ``Scale- and
  context-aware convolutional non-intrusive load monitoring,'' \emph{IEEE
  Transactions on Power Systems}, vol.~35, no.~3, pp. 2362--2373, 2020.

\bibitem{litref06}
M.~U. Hassan, M.~H. Rehmani, R.~Kotagiri, J.~Zhang, and J.~Chen, ``Differential
  privacy for renewable energy resources based smart metering,'' \emph{Journal
  of Parallel and Distributed Computing}, vol. 131, pp. 69--80, 2019.

\bibitem{tkdenew05}
L.~Chen, J.~Zhou, Y.~Chen, Z.~Cao, X.~Dong, and K.-K.~R. Choo, ``Padp:
  Efficient privacy-preserving data aggregation and dynamic pricing for
  vehicle-to-grid networks,'' \emph{IEEE Internet of Things Journal}, vol.~8,
  no.~10, pp. 7863--7873, 2021.

\bibitem{tkdenew15}
U.~Amin, M.~J. Hossain, W.~Tushar, and K.~Mahmud, ``Energy trading in local
  electricity market with renewables—a contract theoretic approach,''
  \emph{IEEE Transactions on Industrial Informatics}, vol.~17, no.~6, pp.
  3717--3730, 2021.

\bibitem{newlit01}
Z.~Almahmoud, J.~Crandall, K.~Elbassioni, T.~T. Nguyen, and M.~Roozbehani,
  ``Dynamic pricing in smart grids under thresholding policies,'' \emph{IEEE
  Transactions on Smart Grid}, vol.~10, no.~3, pp. 3415--3429, 2018.

\bibitem{tkdenew01}
D.~Zhang, H.~Zhu, H.~Zhang, H.~H. Goh, H.~Liu, and T.~Wu, ``Multi objective
  optimization for smart integrated energy system considering demand responses
  and dynamic prices,'' \emph{IEEE Transactions on Smart Grid, in Print}, pp.
  1--1, 2021.

\bibitem{tkdenew13}
A.~S. Khwaja, A.~Anpalagan, M.~Naeem, and B.~Venkatesh, ``Smart meter data
  obfuscation using correlated noise,'' \emph{IEEE Internet of Things Journal},
  vol.~7, no.~8, pp. 7250--7264, 2020.

\bibitem{tkdenew07}
K.~Hafeez, D.~OShea, and M.~H. Rehmani, ``E-dpnct: An enhanced attack resilient
  differential privacy model for smart grids using split noise cancellation,''
  \emph{arXiv preprint arXiv:2110.11091}, 2021.

\bibitem{tkdenew02}
A.~Mohammadali and M.~S. Haghighi, ``A privacy-preserving homomorphic scheme
  with multiple dimensions and fault tolerance for metering data aggregation in
  smart grid,'' \emph{IEEE Transactions on Smart Grid}, vol.~12, no.~6, pp.
  5212--5220, 2021.

\bibitem{myref01}
M.~U. {Hassan}, M.~H. {Rehmani}, and J.~{Chen}, ``Differentially private
  dynamic pricing for efficient demand response in smart grid,'' in \emph{ICC
  2020 - 2020 IEEE International Conference on Communications (ICC)}, 2020, pp.
  1--6.

\bibitem{intref06}
C.~Dwork, ``Differential privacy,'' in \emph{Automata, Languages and
  Programming}.\hskip 1em plus 0.5em minus 0.4em\relax Berlin, Heidelberg:
  Springer Berlin Heidelberg, 2006, pp. 1--12.

\bibitem{addref01}
F.~{Zhao}, X.~{Ren}, S.~{Yang}, Q.~{Han}, P.~{Zhao}, and X.~{Yang}, ``Latent
  dirichlet allocation model training with differential privacy,'' \emph{IEEE
  Transactions on Information Forensics and Security}, vol.~16, pp. 1290--1305,
  2021.

\bibitem{tkderef04}
X.~{Cheng}, P.~{Tang}, S.~{Su}, R.~{Chen}, Z.~{Wu}, and B.~{Zhu}, ``Multi-party
  high-dimensional data publishing under differential privacy,'' \emph{IEEE
  Transactions on Knowledge and Data Engineering}, vol.~32, no.~8, pp.
  1557--1571, 2020.

\bibitem{hassanref01}
M.~U. Hassan, M.~H. Rehmani, and J.~Chen, ``Privacy preservation in blockchain
  based iot systems: Integration issues, prospects, challenges, and future
  research directions,'' \emph{Future Generation Computer Systems}, vol.~97,
  pp. 512--529, 2019.

\bibitem{addref02}
L.~{Wang}, D.~{Zhang}, D.~{Yang}, B.~Y. {Lim}, X.~{Han}, and X.~{Ma}, ``Sparse
  mobile crowdsensing with differential and distortion location privacy,''
  \emph{IEEE Transactions on Information Forensics and Security}, vol.~15, pp.
  2735--2749, 2020.

\bibitem{tkderef06}
T.~{Zhu}, D.~{Ye}, W.~{Wang}, W.~{Zhou}, and P.~{Yu}, ``More than privacy:
  Applying differential privacy in key areas of artificial intelligence,''
  \emph{IEEE Transactions on Knowledge and Data Engineering, in Print}, pp.
  1--1, 2020.

\bibitem{tkderef07}
M.~U. {Hassan}, M.~H. {Rehmani}, and J.~{Chen}, ``Deal: Differentially private
  auction for blockchain-based microgrids energy trading,'' \emph{IEEE
  Transactions on Services Computing}, vol.~13, no.~2, pp. 263--275, 2020.

\bibitem{intref08}
G.~Eibl and D.~Engel, ``Differential privacy for real smart metering data,''
  \emph{Computer Science-Research and Development}, vol.~32, no. 1-2, pp.
  173--182, 2017.

\bibitem{intref07}
M.~U. {Hassan}, M.~H. {Rehmani}, and J.~{Chen}, ``Differential privacy
  techniques for cyber physical systems: A survey,'' \emph{IEEE Communications
  Surveys Tutorials}, vol.~22, no.~1, pp. 746--789, 2020.

\bibitem{newcore01}
M.~Alizadeh, X.~Li, Z.~Wang, A.~Scaglione, and R.~Melton, ``Demand-side
  management in the smart grid: Information processing for the power switch,''
  \emph{IEEE Signal Processing Magazine}, vol.~29, no.~5, pp. 55--67, 2012.

\bibitem{newcore02}
Z.~Fan, P.~Kulkarni, S.~Gormus, C.~Efthymiou, G.~Kalogridis, M.~Sooriyabandara,
  Z.~Zhu, S.~Lambotharan, and W.~H. Chin, ``Smart grid communications: Overview
  of research challenges, solutions, and standardization activities,''
  \emph{IEEE Communications Surveys \& Tutorials}, vol.~15, no.~1, pp. 21--38,
  2012.

\bibitem{tkderef09}
W.~{Liu}, D.~{Qi}, and F.~{Wen}, ``Intraday residential demand response scheme
  based on peer-to-peer energy trading,'' \emph{IEEE Transactions on Industrial
  Informatics}, vol.~16, no.~3, pp. 1823--1835, 2020.

\bibitem{newcore03}
Q.~Shi, C.-F. Chen, A.~Mammoli, and F.~Li, ``Estimating the profile of
  incentive-based demand response (ibdr) by integrating technical models and
  social-behavioral factors,'' \emph{IEEE Transactions on Smart Grid}, vol.~11,
  no.~1, pp. 171--183, 2019.

\bibitem{tkdenew09}
P.~Barbosa, A.~Brito, and H.~Almeida, ``A technique to provide differential
  privacy for appliance usage in smart metering,'' \emph{Information Sciences},
  vol. 370, pp. 355--367, 2016.

\bibitem{tkderef08}
L.~{Ou}, Z.~{Qin}, S.~{Liao}, Y.~{Hong}, and X.~{Jia}, ``Releasing correlated
  trajectories: Towards high utility and optimal differential privacy,''
  \emph{IEEE Transactions on Dependable and Secure Computing}, vol.~17, no.~5,
  pp. 1109--1123, 2020.

\bibitem{algorithmbook}
C.~Dwork, A.~Roth \emph{et~al.}, ``The algorithmic foundations of differential
  privacy.'' \emph{Foundations and Trends in Theoretical Computer Science},
  vol.~9, no. 3-4, pp. 211--407, 2014.

\bibitem{tkdenew10}
S.~M. Ross, \emph{Introduction to probability models}.\hskip 1em plus 0.5em
  minus 0.4em\relax Academic press, 2014.

\bibitem{tkdenew06}
M.~Fire, G.~Katz, L.~Rokach, and Y.~Elovici, ``Links reconstruction attack,''
  in \emph{Security and Privacy in Social Networks}.\hskip 1em plus 0.5em minus
  0.4em\relax Springer, 2013, pp. 181--196.

\bibitem{tkdenew08}
M.~A.~P. Chamikara, P.~Bert{\'o}k, I.~Khalil, D.~Liu, and S.~Camtepe, ``Ppaas:
  Privacy preservation as a service,'' \emph{Computer Communications}, vol.
  173, pp. 192--205, 2021.

\bibitem{expresult}
M.~Muratori, ``Impact of uncoordinated plug-in electric vehicle charging on
  residential power demand-supplementary data,'' National Renewable Energy
  Laboratory-Data (NREL-DATA), Golden, CO (United States), Tech. Rep., 2017.

\bibitem{tkdenew11}
W.~Kong, F.~Luo, Y.~Jia, Z.~Y. Dong, and J.~Liu, ``Benefits of home energy
  storage utilization: An australian case study of demand charge practices in
  residential sector,'' \emph{IEEE Transactions on Smart Grid}, vol.~12, no.~4,
  pp. 3086--3096, 2021.

\end{thebibliography}


\begin{IEEEbiography}[{\includegraphics[width=1in,height=1.25in,clip,keepaspectratio]{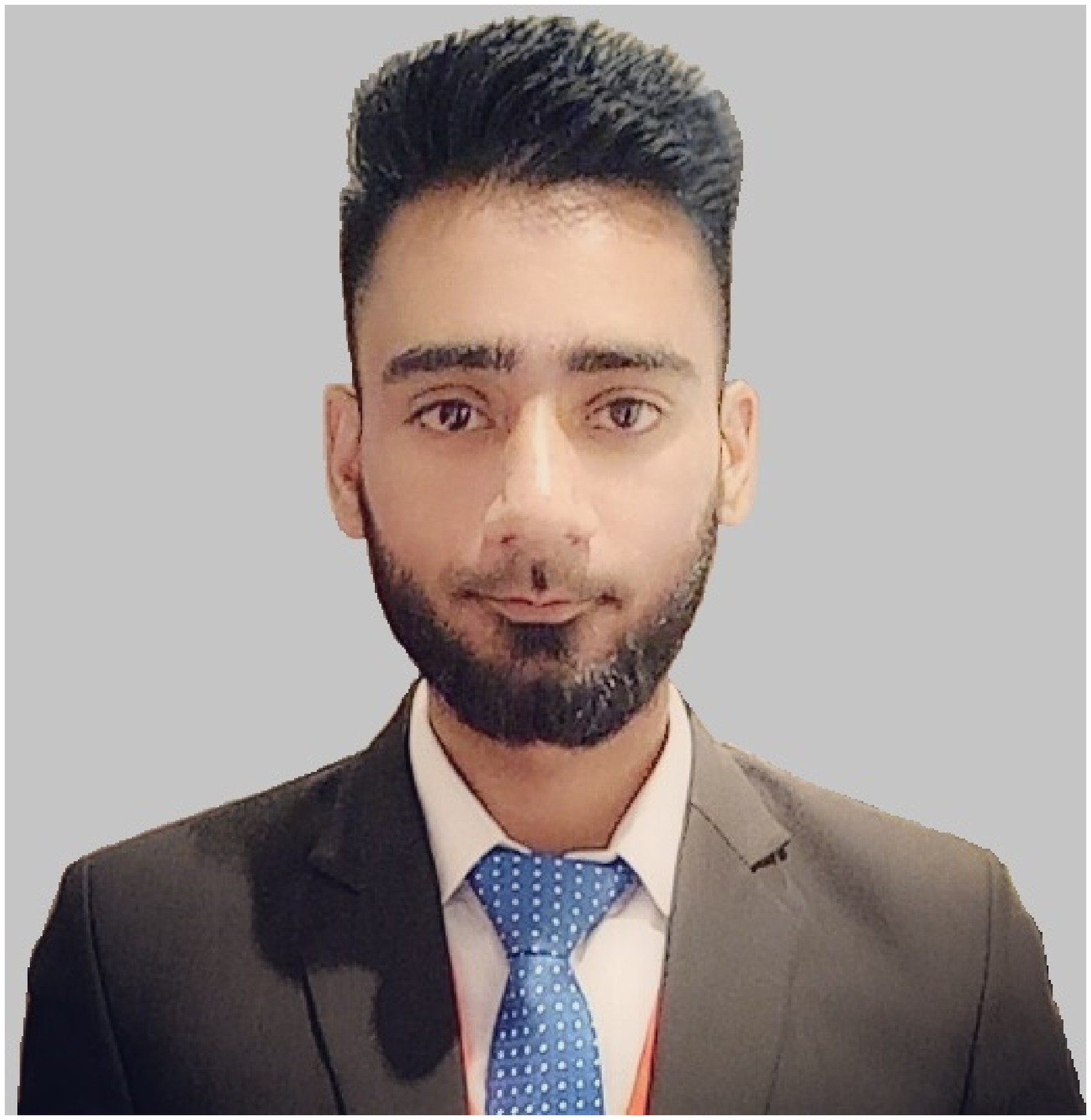}}]{Dr. Muneeb Ul Hassan}

received his PhD degree from Swinburne University of Technology, Australia. He received his Bachelor degree in Electrical Engineering from COMSATS Institute of Information Technology, Wah Cantt, Pakistan, in 2017. He received Gold Medal in Bachelor degree for being topper of Electrical Engineering Department. Currently, he is working as a Research Fellow/ Postdoctoral Researcher at Swinburne University of Technology, Hawthorn VIC 3122, Australia. His research interests include privacy preservation, differential privacy, blockchain, Internet of Things, cyber physical systems, smart grid, cognitive radio networks, and big data.  

\end{IEEEbiography}

\begin{IEEEbiography}[{\includegraphics[width=1in,height=1.25in,clip,keepaspectratio]{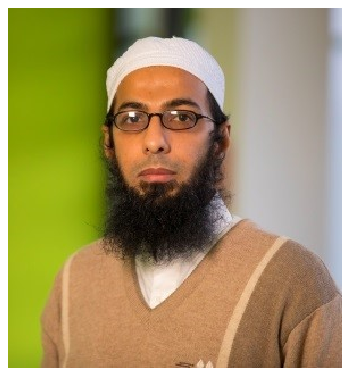}}]{Mubashir Husain Rehmani (M’14-SM’15)}

received the B.Eng. degree in computer systems engineering from Mehran University of Engineering and Technology, Jamshoro, Pakistan, in 2004, the M.S. degree from the University of Paris XI, Paris, France, in 2008, and the Ph.D. degree from the University Pierre and Marie Curie, Paris, in 2011. He is currently working as Assistant Lecturer at Munster Technological University (MTU), Ireland. He received several best paper awards. He is serving in the editorial board of several top ranked journals including NATURE Scientific Reports, IEEE Communication Surveys and Tutorials, IEEE Transactions on Green Communication and Networking and many others. He has been selected for inclusion on the annual Highly Cited Researchers™ 2020 and 2021 list from Clarivate. His performance in this context features in the top 1\% in the field of Computer Science and Cross Field.  

\end{IEEEbiography}

\begin{IEEEbiography}[{\includegraphics[width=1in,height=1.25in,clip,keepaspectratio]{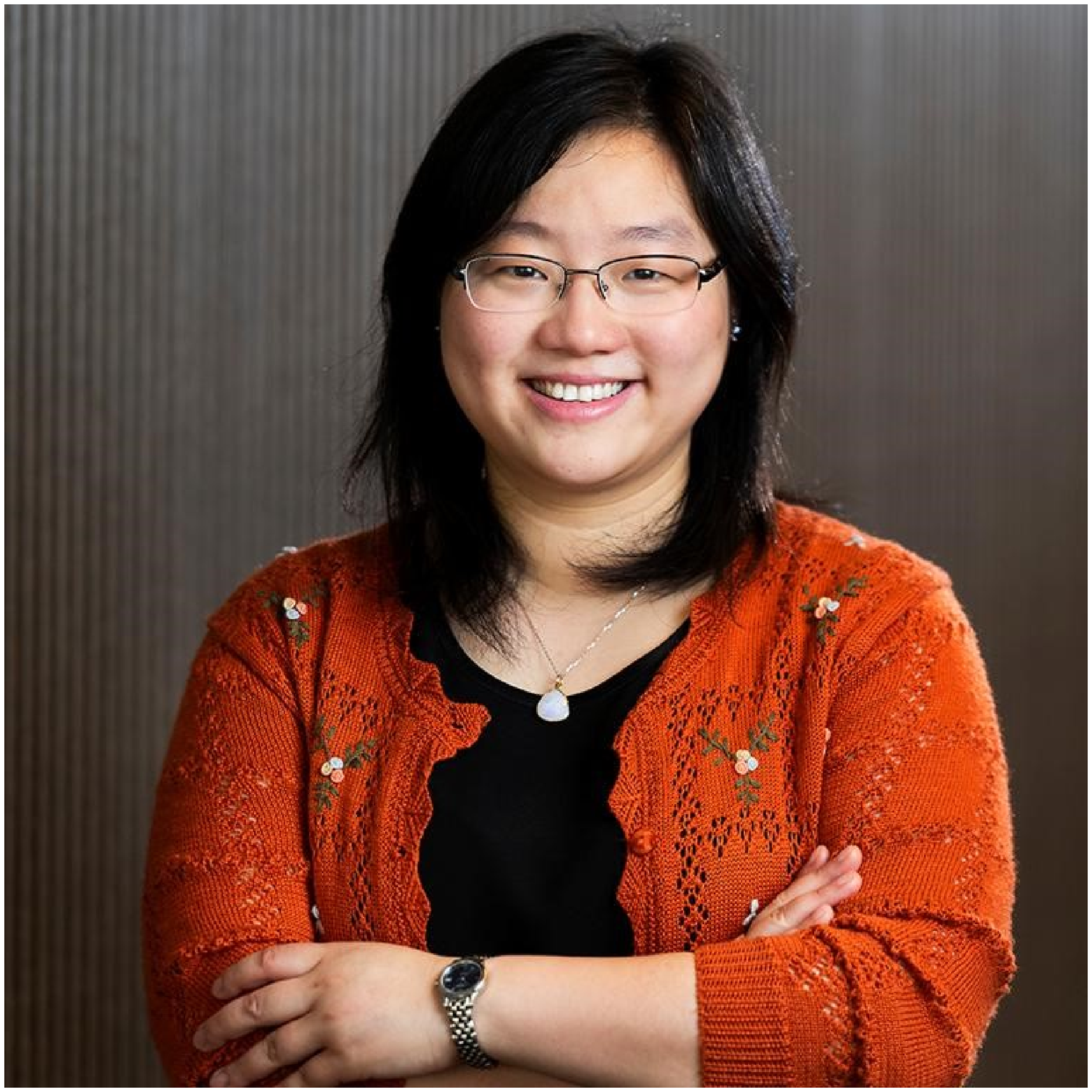}}]{Dr. Jia Tina Du}

is an Associate Professor at The University of South Australia. She received her Ph.D. in Information Science from Queensland University of Technology, Australia. Her research interests lie in information behaviour, interactive information retrieval, and social informatics. 

\end{IEEEbiography}

\begin{IEEEbiography}[{\includegraphics[width=1in,height=1.25in,clip,keepaspectratio]{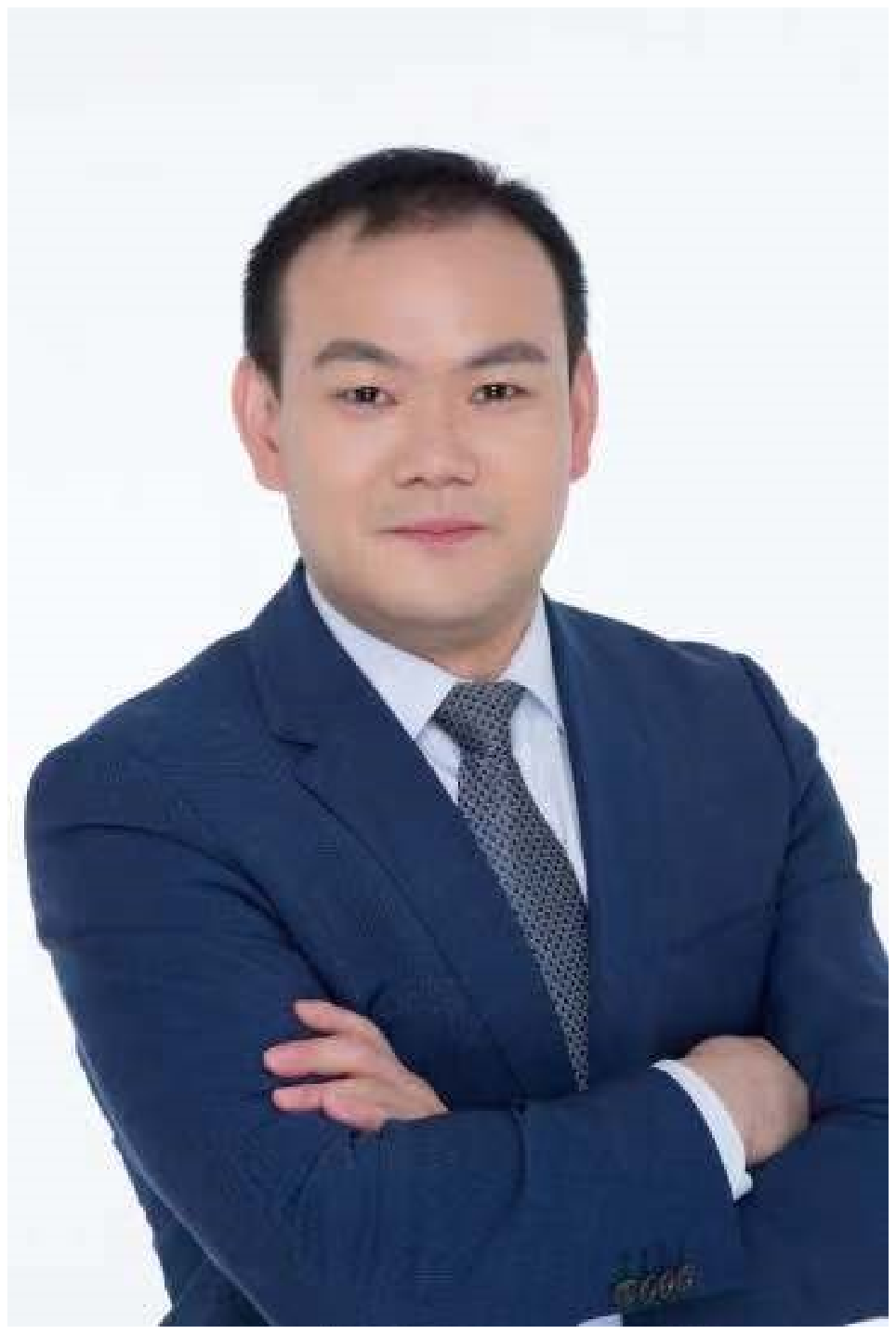}}]{Dr. Jinjun Chen}

is a Professor from Swinburne University of Technology, Australia. He is Deputy Director of Swinburne Data Science Research Institute. He holds a PhD in Information Technology from Swinburne University of Technology, Australia. His research interests include scalability, big data, data science, data systems, cloud computing, data privacy and security, health data analytics and related various research topics. His research results have been published in more than 160 papers in international journals and conferences, including various IEEE/ACM Transactions. 

\end{IEEEbiography}

\end{document}